\input harvmac
%

\let\includefigures=\iftrue
\let\useblackboard=\iftrue
\newfam\black

\includefigures
\message{If you do not have epsf.tex (to include figures),}
\message{change the option at the top of the tex file.}
\input epsf
\def\figin{\epsfcheck\figin}\def\figins{\epsfcheck\figins}
\def\epsfcheck{\ifx\epsfbox\UnDeFiNeD
\message{(NO epsf.tex, FIGURES WILL BE IGNORED)}
\gdef\figin##1{\vskip2in}\gdef\figins##1{\hskip.5in}
\else\message{(FIGURES WILL BE INCLUDED)}%
\gdef\figin##1{##1}\gdef\figins##1{##1}\fi}
\def\DefWarn#1{}
\def\figinsert{\goodbreak\midinsert}
\def\ifig#1#2#3{\DefWarn#1\xdef#1{fig.~\the\figno}
\writedef{#1\leftbracket fig.\noexpand~\the\figno}%
\figinsert\figin{\centerline{#3}}\medskip\centerline{\vbox{
\baselineskip12pt\advance\hsize by -1truein
\noindent\footnotefont{\bf Fig.~\the\figno:} #2}}
\endinsert\global\advance\figno by1}
\else
\def\ifig#1#2#3{\xdef#1{fig.~\the\figno}
\writedef{#1\leftbracket fig.\noexpand~\the\figno}%
\global\advance\figno by1} \fi

\def\journal#1&#2(#3){\unskip, \sl #1\ \bf #2 \rm(19#3) }
\def\andjournal#1&#2(#3){\sl #1~\bf #2 \rm (19#3) }

\def\ie{{\it i.e.}}

\noblackbox
%


\def\unlockat{\catcode`\@=11}
\def\lockat{\catcode`\@=12}

\unlockat

\def\newsec#1{\global\advance\secno by1\message{(\the\secno. #1)}
\global\subsecno=0\global\subsubsecno=0\eqnres@t\noindent
{\bf\the\secno. #1}
\writetoca{{\secsym} {#1}}\par\nobreak\medskip\nobreak}
\global\newcount\subsecno \global\subsecno=0
\def\subsec#1{\global\advance\subsecno
by1\message{(\secsym\the\subsecno. #1)}
\ifnum\lastpenalty>9000\else\bigbreak\fi\global\subsubsecno=0
\noindent{\it\secsym\the\subsecno. #1}
\writetoca{\string\quad {\secsym\the\subsecno.} {#1}}
\par\nobreak\medskip\nobreak}
\global\newcount\subsubsecno \global\subsubsecno=0
\def\subsubsec#1{\global\advance\subsubsecno by1
\message{(\secsym\the\subsecno.\the\subsubsecno. #1)}
\ifnum\lastpenalty>9000\else\bigbreak\fi
\noindent\quad{\secsym\the\subsecno.\the\subsubsecno.}{#1}
\writetoca{\string\qquad{\secsym\the\subsecno.\the\subsubsecno.}{#1}}
\par\nobreak\medskip\nobreak}

\def\subsubseclab#1{\DefWarn#1\xdef
#1{\noexpand\hyperref{}{subsubsection}%
{\secsym\the\subsecno.\the\subsubsecno}%
{\secsym\the\subsecno.\the\subsubsecno}}%
\writedef{#1\leftbracket#1}\wrlabeL{#1=#1}}
\lockat

\def\ie{{\it i.e.}}


\font\manual=manfnt \def\dbend{\lower3.5pt\hbox{\manual\char127}}

\def\IZ{\relax\ifmmode\mathchoice
{\hbox{\cmss Z\kern-.4em Z}}{\hbox{\cmss Z\kern-.4em Z}}
{\lower.9pt\hbox{\cmsss Z\kern-.4em Z}}
{\lower1.2pt\hbox{\cmsss Z\kern-.4em Z}}\else{\cmss Z\kern-.4em
Z}\fi}
\def\half{{1\over 2}}


\def\IZ{\relax\ifmmode\mathchoice
{\hbox{\cmss Z\kern-.4em Z}}{\hbox{\cmss Z\kern-.4em Z}}
{\lower.9pt\hbox{\cmsss Z\kern-.4em Z}}
{\lower1.2pt\hbox{\cmsss Z\kern-.4em Z}}\else{\cmss Z\kern-.4em
Z}\fi}
\def\IB{\relax{\rm I\kern-.18em B}}
\def\IC{{\relax\hbox{$\inbar\kern-.3em{\rm C}$}}}
\def\ID{\relax{\rm I\kern-.18em D}}
\def\IE{\relax{\rm I\kern-.18em E}}
\def\IF{\relax{\rm I\kern-.18em F}}
\def\IG{\relax\hbox{$\inbar\kern-.3em{\rm G}$}}
\def\IGa{\relax\hbox{${\rm I}\kern-.18em\Gamma$}}
\def\IH{\relax{\rm I\kern-.18em H}}
\def\II{\relax{\rm I\kern-.18em I}}
\def\IK{\relax{\rm I\kern-.18em K}}
\def\IP{\relax{\rm I\kern-.18em P}}
\def\IQ{\relax\hbox{$\inbar\kern-.3em{\rm Q}$}}

\def\inbar{\,\vrule height1.5ex width.4pt depth0pt}

\font\cmss=cmss10 \font\cmsss=cmss10 at 7pt
\def\IR{\relax{\rm I\kern-.18em R}}

%
%

\def\makeblankbox#1#2{\hbox{\lower\dp0\vbox{\hidehrule{#1}{#2}%
   \kern -#1
   \hbox to \wd0{\hidevrule{#1}{#2}%
      \raise\ht0\vbox to #1{}
      \lower\dp0\vtop to #1{}
      \hfil\hidevrule{#2}{#1}}%
   \kern-#1\hidehrule{#2}{#1}}}%
}%
\def\hidehrule#1#2{\kern-#1\hrule height#1 depth#2 \kern-#2}%
\def\hidevrule#1#2{\kern-#1{\dimen0=#1\advance\dimen0 by #2\vrule
    width\dimen0}\kern-#2}%
\def\openbox{\ht0=1.2mm \dp0=1.2mm \wd0=2.4mm  \raise 2.75pt
\makeblankbox {.25pt} {.25pt}  }

\def\bun#1/#2{\leavevmode
   \kern.1em \raise .5ex \hbox{\the\scriptfont0 #1}%
   \kern-.1em $/$%
   \kern-.15em \lower .25ex \hbox{\the\scriptfont0 #2}%
}

\def\opensquare{\ht0=3.4mm \dp0=3.4mm \wd0=6.8mm  \raise 2.7pt
\makeblankbox {.25pt} {.25pt}  }


\def\sector#1#2{\ {\scriptstyle #1}\hskip 1mm
\mathop{\opensquare}\limits_{\lower 1mm\hbox{$\scriptstyle#2$}}\hskip 1mm}

\def\tsector#1#2{\ {\scriptstyle #1}\hskip 1mm
\mathop{\opensquare}\limits_{\lower 1mm\hbox{$\scriptstyle#2$}}^\sim\hskip 1mm}


\def\inbar{\,\vrule height1.5ex width.4pt depth0pt}

\font\cmss=cmss10 \font\cmsss=cmss10 at 7pt
\def\IR{\relax{\rm I\kern-.18em R}}


\def\frac#1#2{{#1\over#2}}

\def\half{\frac12}

\def\inbar{\,\vrule height1.5ex width.4pt depth0pt}
\def\IC{\relax\hbox{$\inbar\kern-.3em{\rm C}$}}
\def\IR{\relax{\rm I\kern-.18em R}}
\def\IP{\relax{\rm I\kern-.18em P}}

%
%
\catcode`\@=11
\def\slash#1{\mathord{\mathpalette\c@ncel{#1}}}
\overfullrule=0pt

\def\II{{\cal I}}

\def\underrel#1\over#2{\mathrel{\mathop{\kern\z@#1}\limits_{#2}}}

\catcode`\@=12


%

\def\exp{{\rm exp}}



\def\frac#1#2{{#1\over#2}}

\def\half{\frac12}

\def\inbar{\,\vrule height1.5ex width.4pt depth0pt}
\def\IC{\relax\hbox{$\inbar\kern-.3em{\rm C}$}}
\def\IR{\relax{\rm I\kern-.18em R}}
\def\IP{\relax{\rm I\kern-.18em P}}

%
%

%
\catcode`\@=11
\def\slash#1{\mathord{\mathpalette\c@ncel{#1}}}
\overfullrule=0pt

\def\II{{\cal I}}

\def\underrel#1\over#2{\mathrel{\mathop{\kern\z@#1}\limits_{#2}}}

\catcode`\@=12


%

\def\exp{{\rm exp}}



\lref\GiveonSR{
  A.~Giveon and D.~Kutasov,
  ``Brane dynamics and gauge theory,''
  Rev.\ Mod.\ Phys.\  {\bf 71}, 983 (1999)
  [arXiv:hep-th/9802067].
}

\lref\ElitzurFH{
  S.~Elitzur, A.~Giveon and D.~Kutasov,
  ``Branes and N = 1 duality in string theory,''
  Phys.\ Lett.\ B {\bf 400}, 269 (1997)
  [arXiv:hep-th/9702014].
}

\lref\ElitzurHC{
  S.~Elitzur, A.~Giveon, D.~Kutasov, E.~Rabinovici and A.~Schwimmer,
  ``Brane dynamics and N = 1 supersymmetric gauge theory,''
  Nucl.\ Phys.\ B {\bf 505}, 202 (1997)
  [arXiv:hep-th/9704104].
}

\lref\HananyIE{
A.~Hanany and E.~Witten,
``Type IIB superstrings, BPS monopoles, and three-dimensional gauge
dynamics,''
Nucl.\ Phys.\  B {\bf 492}, 152 (1997)
[arXiv:hep-th/9611230].
}

\lref\SeibergPQ{
  N.~Seiberg,
  ``Electric - magnetic duality in supersymmetric non - Abelian gauge theories,''
  Nucl.\ Phys.\  B {\bf 435}, 129 (1995)
  [arXiv:hep-th/9411149].
}

\lref\GiveonEF{
  A.~Giveon and D.~Kutasov,
  ``Stable and Metastable Vacua in SQCD,''
  arXiv:0710.0894 [hep-th].
}

\lref\OoguriBG{
  H.~Ooguri and Y.~Ookouchi,
  ``Meta-stable supersymmetry breaking vacua on intersecting branes,''
  Phys.\ Lett.\ B {\bf 641}, 323 (2006)
  [arXiv:hep-th/0607183].
}

\lref\FrancoHT{
  S.~Franco, I.~Garcia-Etxebarria and A.~M.~Uranga,
  ``Non-supersymmetric meta-stable vacua from brane configurations,''
  JHEP {\bf 0701}, 085 (2007)
  [arXiv:hep-th/0607218].
}

\lref\BenaRG{
  I.~Bena, E.~Gorbatov, S.~Hellerman, N.~Seiberg and D.~Shih,
  ``A note on (meta)stable brane configurations in MQCD,''
  JHEP {\bf 0611}, 088 (2006)
  [arXiv:hep-th/0608157].
}

\lref\AhnGN{
  C.~Ahn,
  ``Brane configurations for nonsupersymmetric meta-stable vacua in SQCD with
  adjoint matter,''
  Class.\ Quant.\ Grav.\  {\bf 24}, 1359 (2007)
  [arXiv:hep-th/0608160].
}

\lref\AhnTG{
  C.~Ahn,
  ``M-theory lift of meta-stable brane configuration in symplectic and
  orthogonal gauge groups,''
  Phys.\ Lett.\  B {\bf 647}, 493 (2007)
  [arXiv:hep-th/0610025].
}

\lref\ArgurioNY{
  R.~Argurio, M.~Bertolini, S.~Franco and S.~Kachru,
  ``Gauge/gravity duality and meta-stable dynamical supersymmetry breaking,''
  JHEP {\bf 0701}, 083 (2007)
  [arXiv:hep-th/0610212].
}

\lref\AganagicEX{
  M.~Aganagic, C.~Beem, J.~Seo and C.~Vafa,
  ``Geometrically induced metastability and holography,''
  arXiv:hep-th/0610249.
}

\lref\TatarDM{
  R.~Tatar and B.~Wetenhall,
  ``Metastable vacua, geometrical engineering and MQCD transitions,''
  JHEP {\bf 0702}, 020 (2007)
  [arXiv:hep-th/0611303].
}

\lref\KitanoXG{
  R.~Kitano, H.~Ooguri and Y.~Ookouchi,
  ``Direct mediation of meta-stable supersymmetry breaking,''
  Phys.\ Rev.\  D {\bf 75}, 045022 (2007)
  [arXiv:hep-ph/0612139].
}

\lref\HeckmanWK{
  J.~J.~Heckman, J.~Seo and C.~Vafa,
  ``Phase Structure of a Brane/Anti-Brane System at Large N,''
  JHEP {\bf 0707}, 073 (2007)
  [arXiv:hep-th/0702077].
}

\lref\GiveonFK{
  A.~Giveon and D.~Kutasov,
  ``Gauge symmetry and supersymmetry breaking from intersecting branes,''
  Nucl.\ Phys.\  B {\bf 778}, 129 (2007)
  [arXiv:hep-th/0703135].
}

\lref\ArgurioQK{
  R.~Argurio, M.~Bertolini, S.~Franco and S.~Kachru,
  ``Metastable vacua and D-branes at the conifold,''
  JHEP {\bf 0706}, 017 (2007)
  [arXiv:hep-th/0703236].
}

\lref\MurthyQM{
  S.~Murthy,
  ``On supersymmetry breaking in string theory from gauge theory in a throat,''
  arXiv:hep-th/0703237.
}

\lref\DouglasTU{
  M.~R.~Douglas, J.~Shelton and G.~Torroba,
  ``Warping and supersymmetry breaking,''
  arXiv:0704.4001 [hep-th].
}

\lref\MarsanoFE{
  J.~Marsano, K.~Papadodimas and M.~Shigemori,
  ``Nonsupersymmetric brane / antibrane configurations in type IIA and M
  theory,''
  arXiv:0705.0983 [hep-th].
}

\lref\HeckmanUB{
  J.~J.~Heckman and C.~Vafa,
  ``Geometrically Induced Phase Transitions at Large N,''
  arXiv:0707.4011 [hep-th].
}

\lref\AganagicKD{
  M.~Aganagic, C.~Beem and B.~Freivogel,
  ``Geometric Metastability, Quivers and Holography,''
  arXiv:0708.0596 [hep-th].
}

\lref\MazzucatoAH{
  L.~Mazzucato, Y.~Oz and S.~Yankielowicz,
  ``Supersymmetry Breaking Vacua from M Theory Fivebranes,''
  arXiv:0709.2491 [hep-th].
}

\lref\AganagicPY{
  M.~Aganagic, C.~Beem and S.~Kachru,
  ``Geometric Transitions and Dynamical SUSY Breaking,''
  arXiv:0709.4277 [hep-th].
}

\lref\AhnGB{
  C.~Ahn,
  ``Meta-Stable Brane Configurations of Multiple Product Gauge Groups with
  arXiv:0710.0180 [hep-th].
}

\lref\BarbonZU{
  J.~L.~F.~Barbon,
  ``Rotated branes and N = 1 duality,''
  Phys.\ Lett.\  B {\bf 402}, 59 (1997)
  [arXiv:hep-th/9703051].
}

\lref\IntriligatorDD{
  K.~Intriligator, N.~Seiberg and D.~Shih,
  ``Dynamical SUSY breaking in meta-stable vacua,''
  JHEP {\bf 0604}, 021 (2006)
  [arXiv:hep-th/0602239].
}

\lref\AmaritiVK{
  A.~Amariti, L.~Girardello and A.~Mariotti,
  ``Non-supersymmetric meta-stable vacua in SU(N) SQCD with adjoint matter,''
  JHEP {\bf 0612}, 058 (2006)
  [arXiv:hep-th/0608063].
}

\lref\CallanAT{
  C.~G.~Callan, J.~A.~Harvey and A.~Strominger,
  ``Supersymmetric string solitons,''
  arXiv:hep-th/9112030.
}

\lref\HabaRJ{
N.~Haba and N.~Maru,
``A Simple Model of Direct Gauge Mediation of Metastable Supersymmetry
Breaking,''
arXiv:0709.2945 [hep-ph].
}

\lref\IntriligatorPY{
  K.~Intriligator, N.~Seiberg and D.~Shih,
  ``Supersymmetry Breaking, R-Symmetry Breaking and Metastable Vacua,''
  JHEP {\bf 0707}, 017 (2007)
  [arXiv:hep-th/0703281].
}

\lref\SchmaltzSQ{
  M.~Schmaltz and R.~Sundrum,
  ``N = 1 field theory duality from M-theory,''
  Phys.\ Rev.\  D {\bf 57}, 6455 (1998)
  [arXiv:hep-th/9708015].
}

\lref\HoriIW{
  K.~Hori,
  ``Branes and electric-magnetic duality in supersymmetric {QCD},''
  Nucl.\ Phys.\  B {\bf 540}, 187 (1999)
  [arXiv:hep-th/9805142].
}

\Title{
} {\vbox{ \centerline{Stable and Metastable Vacua}
\bigskip
\centerline{in Brane Constructions of SQCD} }}
\medskip
\centerline{\it Amit Giveon${}^{1}$ and David Kutasov${}^{2}$}
\bigskip
\smallskip
\centerline{${}^{1}$Racah Institute of Physics, The Hebrew
University} \centerline{Jerusalem 91904, Israel}
\smallskip
\centerline{${}^2$EFI and Department of Physics, University of
Chicago} \centerline{5640 S. Ellis Av., Chicago, IL 60637, USA }

\bigskip\bigskip\bigskip
\noindent
In a recent paper \GiveonEF\ we showed that $N=1$ supersymmetric QCD in
the presence of certain superpotential deformations has a rich landscape
of supersymmetric and non-supersymmetric vacua. In this paper we
embed this theory in string theory as a low energy theory of intersecting
NS and D-branes. We find that in the region of parameter space of brane
configurations that can be reliably studied using classical string theory,
the vacuum structure is qualitatively similar to that in the field theory regime.
Effects that in field theory come from one loop corrections arise  in
string theory as classical gravitational effects. The brane construction
provides a  useful  guide to the  structure of stable and metastable gauge
theory vacua.

\vglue .3cm
\bigskip

\Date{10/07}

\bigskip

\newsec{Introduction}

Systems of intersecting Neveu-Schwarz (NS) fivebranes and D-branes provide a natural
embedding of field theories into string theory. Many non-trivial aspects of the
vacuum structure of various field theories appear naturally in these brane constructions.
Conversely, field theoretic dynamics sheds light on the behavior of the branes in
regions of parameter space where they are difficult to study using other means.

Most of the original work on this subject focused on theories with unbroken 
supersymmetry (see \GiveonSR\ for a review). More recently, interest turned 
to non-supersymmetric dynamics. In \IntriligatorDD\ it was found that $N=1$ 
supersymmetric QCD (SQCD) has metastable vacua which can be reliably studied
using field theoretic techniques in certain regions of coupling space. The 
brane realization of these vacua and many generalizations were discussed in
\refs{\OoguriBG\FrancoHT\BenaRG\AmaritiVK\AhnGN\AhnTG\ArgurioNY\AganagicEX\TatarDM\KitanoXG
\HeckmanWK\GiveonFK\ArgurioQK\MurthyQM\DouglasTU\MarsanoFE\HeckmanUB\AganagicKD
\MazzucatoAH\AganagicPY-\AhnGB}.
As is familiar from other contexts, the
region in the parameter space of brane configurations which can be reliably studied
using classical string theory is different from the one in which the gauge theory
analysis of \IntriligatorDD\ is valid. Nevertheless, it was found in
\refs{\OoguriBG\FrancoHT-\BenaRG,\GiveonFK} that the pattern
of metastable vacua in the brane systems is very similar to the gauge theory one,
although the detailed dynamics is different.

In particular, in the field theoretic analysis of \IntriligatorDD\  one loop corrections to
the potential for the light fields play an important role in giving mass to certain fields
which classically have an exactly flat potential. It was pointed out in \GiveonFK\ that
in the regime of validity of classical string theory this mass is due to gravitational
attraction of the D-branes to the $NS5$-branes.
This phenomenon is reminiscent of worldsheet duality, whereby the interaction between
D-branes can be studied by calculating the one loop partition sum of light open strings
when the D-branes are close, and by taking into account the exchange of light closed
strings when they are far apart.  As there,  the two regimes are qualitatively similar.

In this paper we continue our investigation of supersymmetry breaking  in intersecting
brane systems, by generalizing the analysis of \GiveonFK\ to a configuration with a richer
vacuum structure. On the field theory side this system is just SQCD with a particular
superpotential perturbation turned on. Its vacuum structure in gauge theory is discussed in
\GiveonEF. On the string theory side, the system we will study is that of
\refs{\OoguriBG\FrancoHT-\BenaRG, \GiveonFK}, with some of the branes rotated.

It was shown in \GiveonEF\ that in the field theory regime this system has a rich landscape 
of supersymmetric and (metastable) non-supersymmeric vacua. Our main purpose here will be to 
demonstrate that this landscape appears in the brane construction as well. As we will see, 
the brane picture provides a good way of identifying and studying both supersymmetric and 
non-supersymmetric ground states, albeit in a different regime in parameter space.

The plan of the paper is as follows. In section 2 we describe the brane configurations
and study  their ground states. In the gauge theory analysis of \GiveonEF, classically
the only ground states are supersymmetric. In the brane picture this is reflected in the 
fact that if we neglect the gravitational potential of the $NS5$-branes, all vacua of the 
brane system are supersymmetric.

In section 3 we include the gravitational potential of the $NS5$-branes and find a rich 
pattern of non-supersymmetric metastable states, in addition to the supersymmetric vacua 
of section 2. This pattern is very similar to that seen in gauge theory in \GiveonEF. In 
section 4 we comment on our results.

\newsec{Brane configurations and supersymmetric vacua}

To construct the brane configurations that we will study, it is convenient to decompose
the $9+1$ dimensional spacetime as follows:
\eqn\decomp{\IR^{9,1}=\IR^{3,1}\times \IC_v\times \IC_w\times\IR_y\times\IR_{x^7}~.}
The $\IR^{3,1}$ labeled by $(x^0,x^1,x^2,x^3)$ is common to all the branes,
and is the arena for the dynamics of interest. The two complex planes $\IC_v$,
$\IC_w$ and real line $\IR_y$ correspond to
\eqn\nsns{v=x^4+ix^5~,\qquad w=x^8+ix^9~,\qquad y=x^6~.}
We will place in this spacetime  extended branes that intersect on $\IR^{3,1}$,
as well as $D4$-branes localized at the intersection. The extended branes are
$NS5$-branes and $D6$-branes filling $\IR^{3,1}$ and the complex
plane labeled by
\eqn\strns{w_\theta=v\sin\theta+w\cos\theta}
in $\IC_v\times\IC_w$, and localized in the direction transverse
to \strns, $v_\theta=v\cos\theta-w\sin\theta$. We will refer to them as
$NS_\theta$ and $D6_\theta$-branes, respectively. The $D6$-branes are
further stretched in $x^7$.

We will study brane configurations that contain two $NS5$-branes and a
stack of $D6$-branes. In general, all three will have different orientations
in $\IC_v\times\IC_w$ (\ie\ different values of $\theta$, \strns). By choosing the
coordinates $(v,w)$ appropriately, we can take one of the fivebranes to lie
along the $v$ axis. We will do that throughout the
discussion, and refer to the corresponding $NS5$-brane as an $NS$-brane
(following customary notation \GiveonSR). In terms of the definitions above,
this fivebrane corresponds to $NS_{\pi\over2}$. The second fivebrane will be
often taken to be $NS_0$; we will denote it by $NS'$, again following \GiveonSR.

To summarize,  the  extended branes will be taken to stretch in $\IR^{3,1}$ as well
as the following directions:
\eqn\extbranes{\eqalign{NS:&\qquad v~,\cr
                                       NS_{\theta'}:&\qquad w_{\theta'}~,\cr
                                       D6_\theta:&\qquad w_\theta, x^7~.\cr
}}
For general $\theta$, $\theta'$ these branes preserve
$N=1$ supersymmetry in four dimensions. For some special values
of the angles, the supersymmetry is enhanced to $N=2$.

We will also consider $D4$-branes which fill $\IR^{3,1}$ and in the extra dimensions
are stretched between pairs of  the extended branes \extbranes. Adding the
$D4$-branes leads to configurations which may or may not preserve supersymmetry.
If  the direction in which the fourbranes are stretched
is $y$, \nsns, the full brane configuration is supersymmetric \GiveonSR, but
in general it is not.  Our main purpose below will be to analyze the
supersymmetric and non-supersymmetric vacua of brane configurations
containing all the branes listed above, and compare the resulting vacuum
structure to the $3+1$ dimensional effective field theory of the light
modes localized at the intersection.

In the examples we will consider, that theory is SQCD with gauge group $U(N_c)$, 
$N_f$ flavors of chiral superfields in the fundamental representation, $Q_i$, 
$\widetilde Q^i$, and in general a non-zero superpotential $W_{\rm e}(\widetilde Q Q)$. 
We will also consider brane configurations whose low energy dynamics is described by 
the Seiberg dual magnetic gauge theory \SeibergPQ, with gauge group $U(N_f-N_c)$,
$N_f$ flavors of fundamentals $q^i$, $\tilde q_i$, gauge singlet chiral superfields
$M^i_j$, which are magnetic duals of the electric meson fields $\widetilde Q^iQ_j$,
and superpotential
\eqn\wmaggg{W_{\rm mag}={1\over\Lambda}\tilde q_iM^i_jq^j+W_{\rm m}(M)~,}
where $\Lambda$ is  a scale familiar from studies of Seiberg duality. The magnetic
quarks $q$, $\tilde q$ will be taken below to be canonically normalized, while for $M$
it will be convenient to use a different normalization.

The supersymmetric vacuum structure of  brane configurations of the sort described above
was analyzed and matched to field theory some time ago (see \GiveonSR\ for a review). Our
main purpose here is to generalize these results to metastable supersymmetry breaking vacua.
We will see that the brane construction is very useful in identifying and analyzing such states.

\subsec{The magnetic SQCD brane configuration}

We start with the brane configuration of figure 1, whose low energy limit is the magnetic gauge
theory described above with $W_{\rm m}(M)=0$, \refs{\ElitzurFH,\ElitzurHC}.

\ifig\loc{The magnetic brane configuration.}
{\epsfxsize5.0in\epsfbox{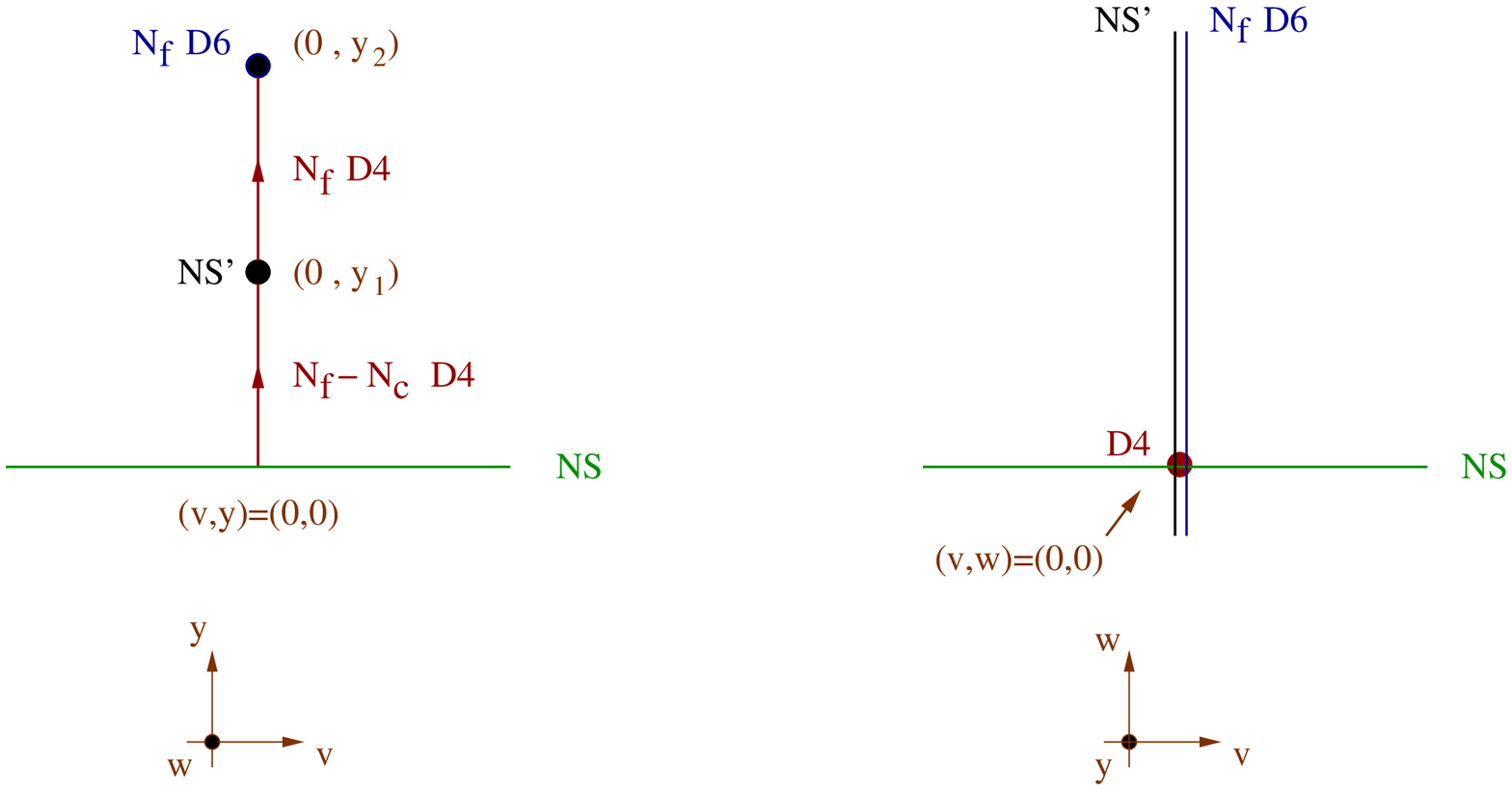}}

\noindent
The figure on the left is a two dimensional slice of the brane configuration. The plane of
the page is labeled by (one of the two components of) $v$ and $y$ (see \nsns). The direction
out of the page is $w$, and the slice is taken at the  location of the $NS$-brane, $w=0$.
The figure on the right is a projection of the brane configuration on $\IC_v\times\IC_w$.
In it, one can think of $y$ as coming out of the page, and the different extended branes are located
at different values of $y$. Comparing to the figure on the left we see that when viewed from
above (in $y$), the order of the branes is:  $D6$-branes followed by the $NS'$-brane and
then the $NS$-brane.\foot{In this and subsequent figures we do not specify the value of
$\theta$ for the $D6$-branes, since it can be read off from the figures. In particular,
in  figure 1 the $D6$-branes have $\theta=0$ (see \strns).}

In addition to the extended branes discussed above, the  configuration of figure 1
contains $D4$-branes localized near the origin in the extra dimensions. The dynamics
of these branes is the focus of our analysis. As reviewed in \GiveonSR, the low energy theory
on the $N_f-N_c$ $D4$-branes stretched between the $NS$ and $NS'$-branes is $N=1$
SYM with gauge group $U(N_f-N_c)$. Strings stretched between these ``color $D4$-branes''
and the $N_f$ ``flavor $D4$-branes'' which connect the $D6$-branes to the $NS'$-brane
give rise to $N_f$ fundamental chiral superfields $q^i$, $\tilde q_i$. Finally, strings both of
whose ends lie on the flavor $D4$-branes give rise to gauge singlet superfields $M^i_j$.
The magnetic quarks $q$, $\tilde q$ and singlets $M$ are coupled via the superpotential
\eqn\wmag{W_{\rm mag}={1\over\Lambda}\tilde q_iM^i_jq^j~.}
Many of the parameters of the brane configuration of figure 1 have an interpretation in the
low energy effective field theory \GiveonSR. In particular, the classical $U(N_f-N_c)$ gauge
coupling, $g_{\rm mag}$, is given by
\eqn\ymm{g_{\rm mag}^2={g_sl_s\over y_1}~,}
where $l_s$ and $g_s$ are the string length and ten dimensional string coupling, respectively,
and $y_1$ is the distance between the $NS5$-branes (see figure 1).

The magnetic superpotential \wmag\ has flat directions corresponding to giving an arbitrary
expectation value to the $N_f\times N_f$ matrix $M^i_j$ while setting $q=\tilde q=0$. This
moduli space is realized geometrically in the brane construction via displacements of the
$D4$-branes stretched between the $D6$-branes and the $NS'$-brane in figure 1 in the
direction $w$, which is common to both types of branes.\foot{In order to exhibit the full $N_f^2$
dimensional moduli space one needs to separate the $D6$-branes in $y$; see figure 29 in
\GiveonSR.}

The precise relation between the displacement of the branes and the expectation value of
$M$ can be read off \wmag. A non-zero expectation value $\langle M_j^j\rangle$
gives rise via \wmag\ to a mass  $\langle M_j^j\rangle/\Lambda$ to $q_j$, $\tilde q^j$.
Geometrically, this mass is due to the stretching of the fundamental string between the $j$'th
flavor brane which is displaced by the amount $w_j$, and the color branes in figure 1.
Therefore, we conclude that  the relation between the two is\foot{Recall that the tension of
the fundamental string is $T=1/2\pi\alpha'=1/2\pi l_s^2$.}
\eqn\mwll{{\langle M_j^j\rangle\over\Lambda}={w_j\over 2\pi l_s^2}~.}
Another deformation of the magnetic gauge theory that will be of interest below
is adding to \wmag\ a linear superpotential,
\eqn\wmagm{W_{\rm mag}={1\over\Lambda}\tilde q_iM^i_jq^j-m\Tr M~,}
which is the magnetic dual of a mass term for the electric quarks $Q$, $\widetilde Q$.
In the brane picture this corresponds to displacing the $D6$-branes relative to the
$NS'$-brane in the $v$ plane. We will normalize $M$ such that displacing the
$D6$-branes to $v=v_2$ corresponds to adding to the superpotential the deformation
\wmagm\ with  \GiveonSR
\eqn\mmm{m=-{v_2\over 2\pi l_s^2}~.}
The last deformation that we will consider corresponds to rotations of the $D6$-branes in the
$(v,w)$ hyperplane. In figure 1 the $D6$-branes are stretched in $w$, and one can ask what
happens if we rotate them by an angle $\theta$ so that they are extended in $w_\theta$, \strns.
This corresponds to adding to the superpotential \wmag\ a mass term for $M$,
\eqn\wmagma{W_{\rm mag}={1\over\Lambda}\tilde q_iM^i_jq^j+{\alpha\over 2}\Tr M^2~,}
with $\alpha$  related to $\theta$ as follows:
\eqn\altt{\alpha\Lambda=\tan\theta~.}
To prove this it is convenient to consider the brane configuration depicted in figure 2.
\ifig\loc{A deformed magnetic configuration.}
{\epsfxsize5.0in\epsfbox{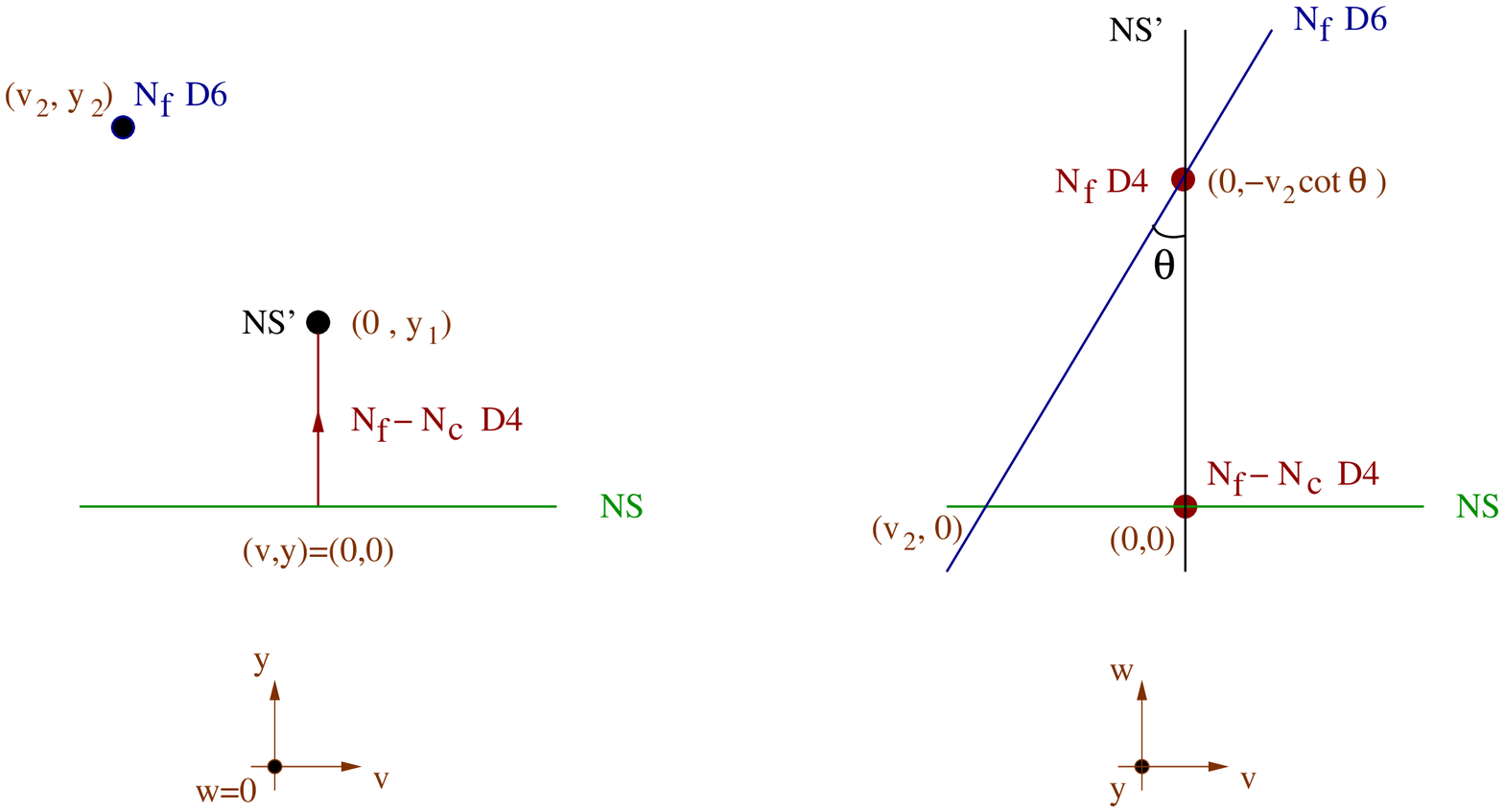}}

\noindent
Since this configuration can be obtained from that of figure 1 by a combination of a translation
of the $D6$-branes in $v$ and a rotation, the corresponding superpotential has both of the deformations
\wmagm\ and \wmagma\ turned on,
\eqn\wmagmb{W_{\rm mag}={1\over\Lambda}\tilde q_iM^i_jq^j+\Tr \left({\alpha\over 2}M^2-mM\right)~.}
The supersymmetric vacuum depicted in figure 2 has $q=\tilde q=0$, so that the gauge group $U(N_f-N_c)$
is unbroken, while the expectation value of $M$ is determined by the F-term condition of \wmagmb:
\eqn\expvalm{M_i^j={m\over\alpha}\delta_i^j~.}
To calculate this expectation value in the brane picture
we use \mwll\ and the value of $w$ for the flavor branes in figure 2:
\eqn\expmm{M_i^j=-{v_2\cot\theta\over2\pi l_s^2}\Lambda\delta_i^j=m\Lambda\delta_i^j\cot\theta~.}
In the second equality we used the relation \mmm. Comparing \expvalm\ and \expmm\ leads to \altt.

The brane configuration of figure 1 has a number of interesting global symmetries. The
$U(N_f)$ gauge symmetry on the $N_f$ coincident $D6$-branes descends in the low
energy gauge theory to the diagonal $U(N_f)$ symmetry of magnetic SQCD. In this
paper we will consider brane configurations that preserve it, as in \GiveonEF. It is easy to
generalize the discussion to configurations that break the $U(N_f)$ symmetry, by
separating the sixbranes. In particular, one can consider generalizations of the
superpotential \wmagmb\ in which $m$ and $\alpha$ are more general matrices in flavor space.

The subgroup of the rotation group of the extra dimensions left unbroken by the
brane configuration of figure 1 is $SO(2)_{45}\times SO(2)_{89}=U(1)_v\times U(1)_w$.
These symmetries are R-symmetries, and in comparing to field theory it is convenient
to normalize the generators such that the supercharges have charge $\pm 1$. The R-charges
of the various fields and parameters can then be deduced from the analysis of deformations above.
The magnetic quarks have charge $(1,0)$, $M$ has charge $(0,2)$, while the couplings
$m$ and $\alpha$ in \wmagmb\ have charge $(2,0)$ and $(2,-2)$, respectively. In particular,
when both couplings are non-zero, the R-symmery is completely broken.

Thus, we see that the brane configuration of figure 2 provides an example of a
background which breaks R-symmetry, and it is of interest to construct
metastable supersymmetry breaking states in it. In the rest of this section we will
describe the supersymmetric vacua of the model. In section 3 we will
turn to metastable states.

Since the brane configuration of figure 2 reduces in the infrared to the magnetic
$U(N_f-N_c)$ gauge theory with the superpotential \wmagmb, one can compare
its vacuum structure to that of the gauge theory.
The gauge theory analysis was done\foot{In \GiveonEF\ the gauge group was taken to be
$SU(N_f-N_c)$; in the string embedding the baryon number symmetry is gauged as well.}
in \GiveonEF, where it was found that classical vacua are labeled by an integer
\eqn\kmag{k=0,1,\dots,N_f-N_c~.}
For given $k$, the expectation values of $M$, $q$ and $\tilde q$ are given by
\eqn\formphi{M=\left(\matrix{0&0\cr 0&{m\over\alpha}I_{N_f-k}\cr}\right)~,}
\eqn\formqq{\tilde qq=\left(\matrix{m\Lambda I_k&0\cr 0&0\cr}\right)~.}
In the $k$'th vacuum the gauge symmetry is broken by the expectation
value of $q$ to $U(N_f-N_c-k)$. Thus, it is clear that the configuration of
figure 2, in which the magnetic gauge group is unbroken, corresponds to
$k=0$.

The remaining vacua are easy to identify in the brane construction as well.
One can take $k$ of the $N_f$ flavor $D4$-branes and connect them to
$k$ of the $N_f-N_c$ color $D4$-branes in figure 2, such that the resulting
branes stretch directly from the $D6$-branes to the $NS$-brane. To minimize
their energy, these $D4$-branes will move to $(v,w)=(v_2,0)$, leading to
the configuration of figure 3.

\ifig\loc{The classical supersymmetric vacua of the deformed magnetic configuration.}
{\epsfxsize5.0in\epsfbox{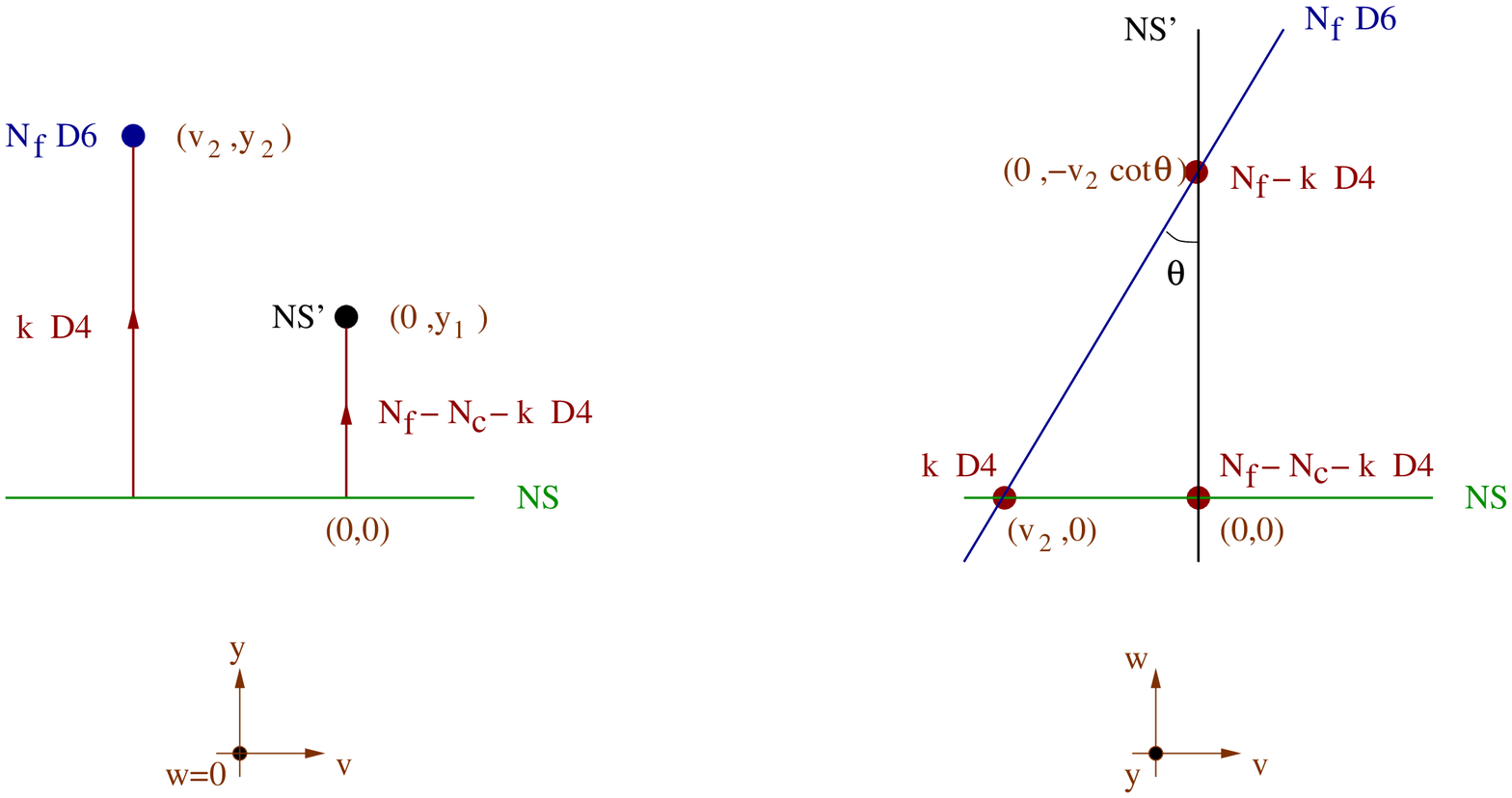}}

\noindent
The $N_f-N_c-k$ $D4$-branes stretched between the $NS5$-branes in figure 3
give the unbroken $U(N_f-N_c-k)$ gauge group. The position in the $w$ plane of
the $N_f-k$ $D4$-branes stretched between the $D6$-branes and $NS'$-brane
is related to the corresponding eigenvalues of $M$ \formphi\ via \mwll.

Thus, we see that the brane analysis reproduces the classical vacuum structure
of magnetic SQCD with the superpotential \wmagmb. Quantum mechanically the
$SU(N_f-N_c-k)$ gauge theory confines and breaks R-symmetry, such that there
are $N_f-N_c-k$ distinct vacua with a mass gap. Furthermore, as shown in \GiveonEF,
for $N_f<2N_c$ there are quantum supersymmetric vacua which are not seen
classically, and are missed by the classical brane construction. This is analogous
to the fact that in the theory with $\alpha=0$ considered in \IntriligatorDD\
there are no classical supersymmetric vacua at all, and the quantum vacua are
not seen (at least naively) in the brane analysis \refs{\OoguriBG\FrancoHT-\BenaRG}.

\subsec{Further deformations}

The brane construction of the previous subsection can be deformed in a number
of ways that do not qualitatively change the low energy physics. One involves the
position of the $D6$-branes in $\IR_y$, $y_2$, and in particular the process of
taking the $D6$-branes past the $NS'$-brane. Starting from the magnetic brane
configuration of figure 3 and continuously changing $y_2$ to a value smaller than
$y_1$  leads to the configuration depicted in figure 4.

\ifig\loc{Another description of the supersymmetric vacua of figure 3.}
{\epsfxsize5.0in\epsfbox{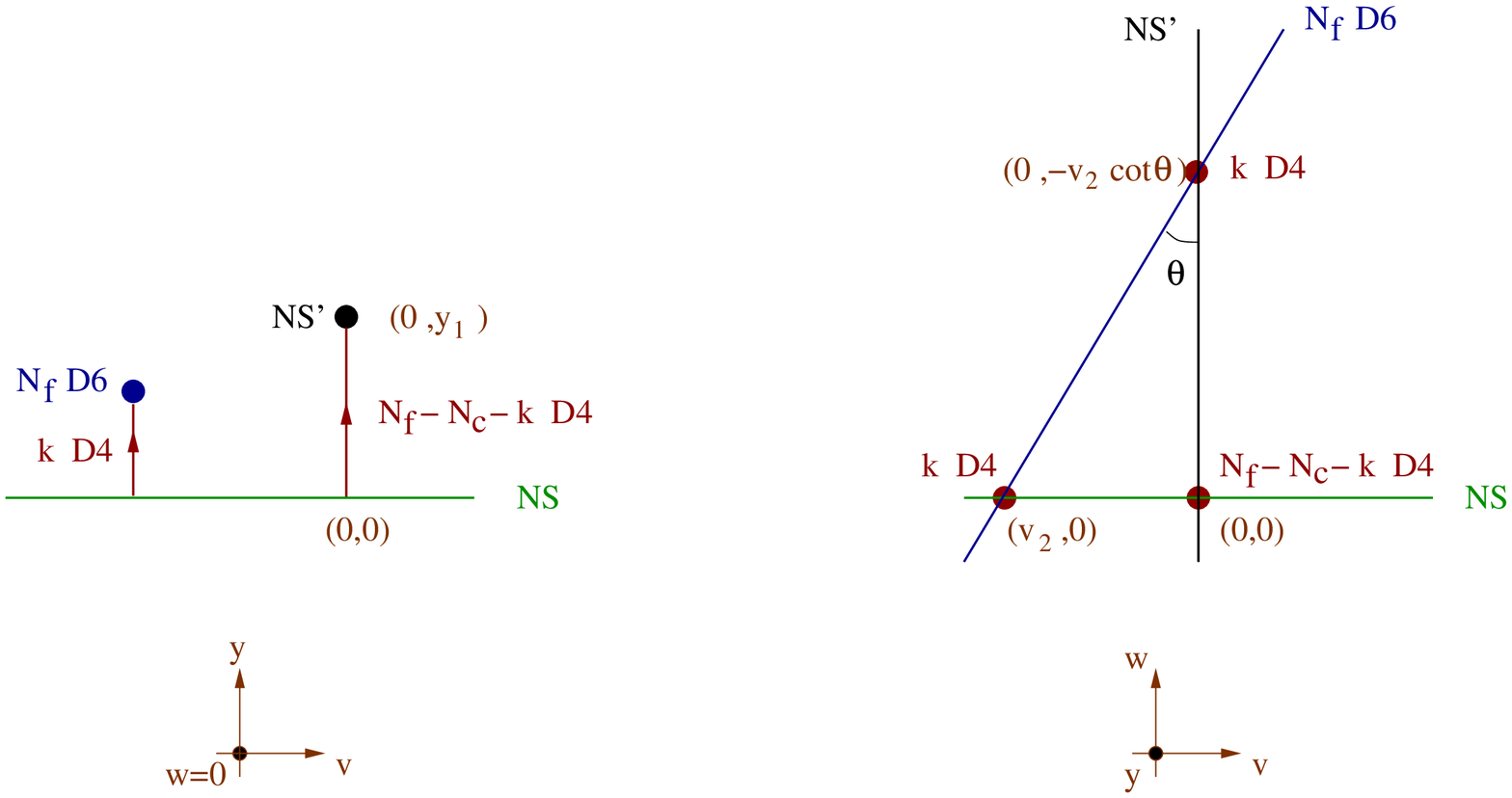}}

\noindent
In arriving at this figure we used the fact that when a $D6$-brane crosses an $NS5$-brane
that is not parallel to it, a $D4$-brane stretching between them is either created or
destroyed via a Hanany-Witten transition \HananyIE.

The low energy effective description of the brane configuration of figure 4 is similar
to that of figure 3, with two differences. One is that it does not include the meson fields
$M$. This is easiest to see by comparing the vacua with $k=0$. In our previous discussion,
the vacuum with $k=0$ is described in figure 2. It includes $N_f$ flavor $D4$-branes stretched
between the $D6$ and $NS'$-branes; the mesons $M_i^j$ are the lowest lying modes of
fundamental strings stretched between these branes. In the gauge theory description they
are massive due to the non-zero value of $\alpha$. In the brane picture  their mass is due to the
non-zero value of the angle $\theta$.

On the other hand, in the configuration of figure 4, for $k=0$ there are no $D4$-branes
stretched between the $D6$ and $NS'$-branes as a result of the Hanany-Witten transition
mentioned above.  Therefore, in this configuration, the meson fields $M$ are absent.
At energies much below the mass of $M$, one can think of the configuration of figure 4 as
obtained from that of figure 3 by integrating this field out, and the behavior of the two
systems is very similar. Above the mass of $M$ they are different (see \GiveonEF\ for further
discussion of this issue).

The second difference between the two brane configurations is of a more quantitative nature.
The effective superpotential of the brane configuration of figure 4 is given by
\eqn\wmagqq{W_{\rm mag}=
\Tr\left[m_q\tilde q q-{\alpha_q\over2}(\tilde q q)^2\right]~,}
which has the same qualitative form as what one would get by integrating out
$M$ in \wmagmb, but the coefficients $m_q$ and $\alpha_q$ are different.

To determine $m_q$, consider the vacuum with $k=0$ in figure 4. In this vacuum
the expectation values of $q$, $\tilde q$ vanish and the $U(N_f-N_c)$ gauge group
is unbroken. As is clear from \wmagqq, the mass of $q$ and $\tilde q$ in this vacuum
is given (up to an unimportant phase) by $m_q$. On the other hand, in figure 4 this mass
corresponds to the energy of the lowest lying fundamental string stretched between the
color $D4$-branes and the $N_f$ $D6$-branes. That energy is given by
$|v_2\cos\theta|/2\pi l_s^2=|m\cos\theta|$ (see \mmm).  Therefore, we conclude that
\eqn\mmqq{m_q=m\cos\theta~.}
Integrating out $M$ from \wmagmb\ gives instead
\eqn\wmagqqqq{W_{\rm mag}=
-{1\over\alpha\Lambda}\Tr\left[{1\over 2\Lambda}(\tilde q q)^2-m\tilde q q\right]~,}
and hence
\eqn\formmmqq{m_q={m\over\alpha\Lambda}=m\cot\theta}
(see \altt). Thus, we see that under the
Hanany-Witten transition, the mass $m_q$ changes by a factor of $\sin\theta$.
Determining $\alpha_q$ is more difficult, but it is easy to see that
$\alpha_q\Lambda$ is a function of $\theta$ which vanishes at  $\theta=0,\pi/2$
and is symmetric under $\theta\to{\pi\over2}-\theta$, which is not a property of
the coefficient of $(\tilde q q)^2$ in \wmagqqqq.

Further displacing the $D6$-branes past the $NS$-brane gives rise to a third brane
configuration that describes the magnetic vacua above. This description can be obtained
from that of subsection 2.1 by exchanging $v$ and $w$ and taking $\theta\to{\pi\over2}-\theta$.
In particular, the coupling $\alpha$ changes to $\alpha\Lambda=\cot\theta$ (compare to \altt).
Thus, we see that Hanany-Witten transitions do not change the qualitative form of the
low energy Lagrangian, but act non-trivially on the couplings.

Another deformation that does not qualitatively affect the low energy physics corresponds to
a rotation of the $NS'$-brane by the angle $\theta'$ into the direction $w_{\theta'}$ \strns.
The resulting configuration is depicted in figure 5, where we introduced the notation
$\theta_1={\pi\over2}+\theta'$,  $\theta_2={\pi\over2}-\theta$.

\ifig\loc{The magnetic configuration with rotated $NS'$-brane.}
{\epsfxsize5.0in\epsfbox{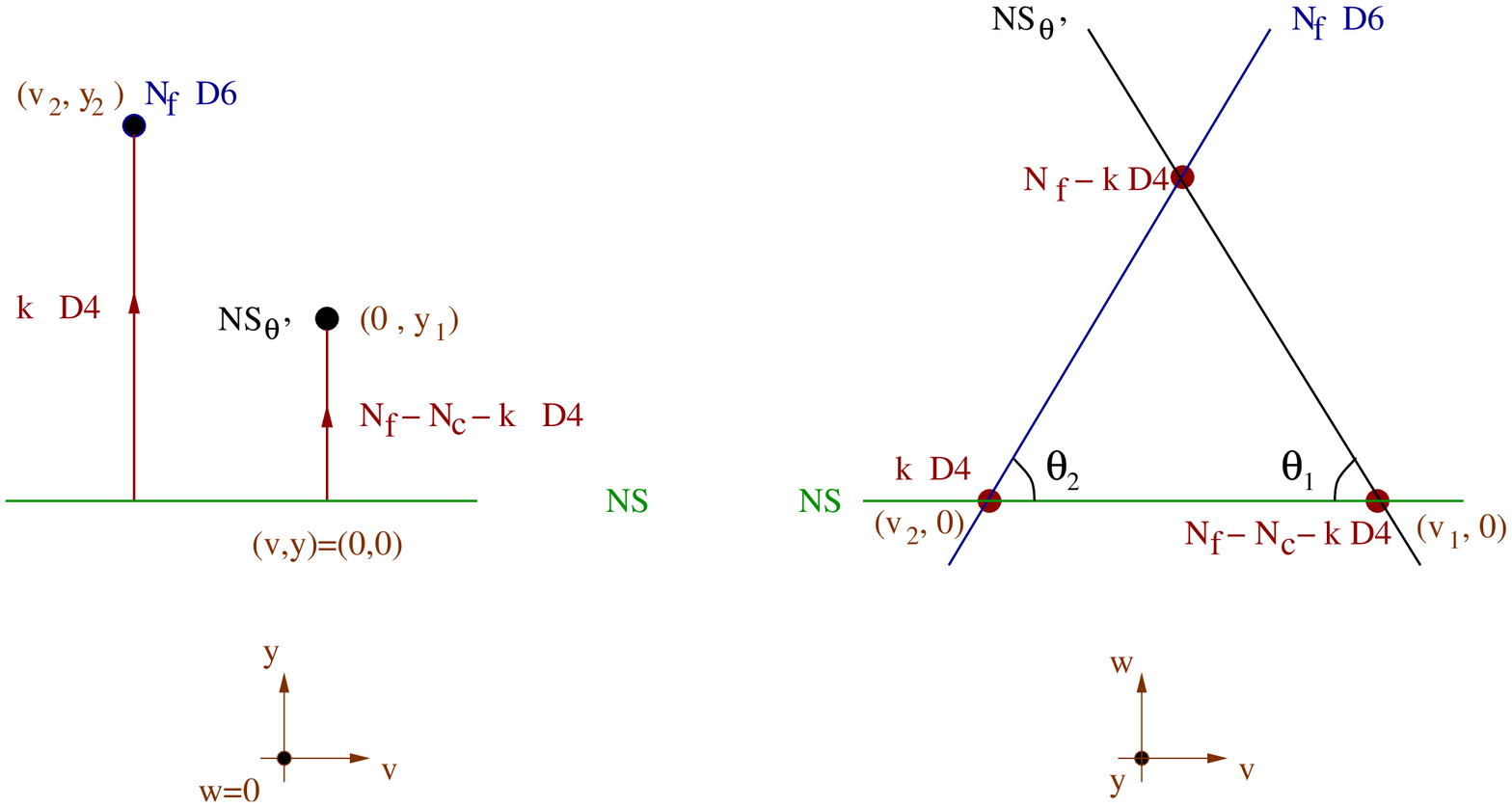}}

\noindent
The rotation of the fivebrane introduces into the dynamics an adjoint of $U(N_f-N_c)$
whose mass goes to infinity in the limit $\theta'\to 0$, but is finite for generic $\theta'$ \BarbonZU.
This field couples to the magnetic quarks via a Yukawa superpotential. Integrating it out leads
to a further contribution to the quartic superpotential for $q$ and $\tilde q$, which goes to
zero as $\theta'\to 0$ (\ie\ $\theta_1\to{\pi\over2}$). For $\theta'=0$ the superpotential is given
by \wmagqqqq; for generic $\theta_1$, $\theta_2$, the superpotential is given by \wmagqq, and
the parameters $m_q$ and $\alpha_q$ can be calculated as before. The former takes the form
\eqn\mvvll{m_q={(v_1-v_2)\sin\theta_2\over 2\pi  l_s^2\sin(\theta_1+\theta_2)}~.}
As $\theta_1\to{\pi\over2}$ one recovers our previous result \formmmqq.

Note that $m_q$ \mvvll\ diverges as $\theta_1+\theta_2\to\pi$. In this limit, the $NS_{\theta'}$
and $D6$-branes become parallel which implies that the coefficient of $M^2$ in the superpotential
goes to zero. The effective superpotential of $M$ and $q$ in this brane configuration has
the form
\eqn\neweffsup{W_{\rm mag}={1\over\Lambda} \tilde qM q-mM+\beta(\tilde q q)^2~.}
The last term can be thought of as due to integrating out the adjoint of $U(N_f-N_c)$
discussed above. The coupling $\beta$ depends on $\theta_1$ (or equivalently on
$\theta_2=\pi-\theta_1$), and goes to zero as $\theta_1,\theta_2\to\pi/2$.  This is precisely
the model that was studied recently in \HabaRJ\ in the context of direct gauge mediation.
Like the original model of \IntriligatorDD, it does not have classical supersymmetric vacua,
as is clear from the brane construction. Nevertheless, for generic $\theta_1=\pi-\theta_2$
it breaks the R-symmetry. In the gauge theory this is due to the extra term in the superpotential
\neweffsup, while in the brane construction it is clear from the geometry that $U(1)_v\times
U(1)_w$ is  broken  in figure 5.

\subsec{The electric configuration}

The brane configurations discussed above give rise at low energies to magnetic
SQCD, which is related by Seiberg duality \SeibergPQ\  to an electric  theory with
gauge group $U(N_c)$. The corresponding brane configuration is presented in figure 6.

\ifig\loc{The electric configuration.}
{\epsfxsize5.0in\epsfbox{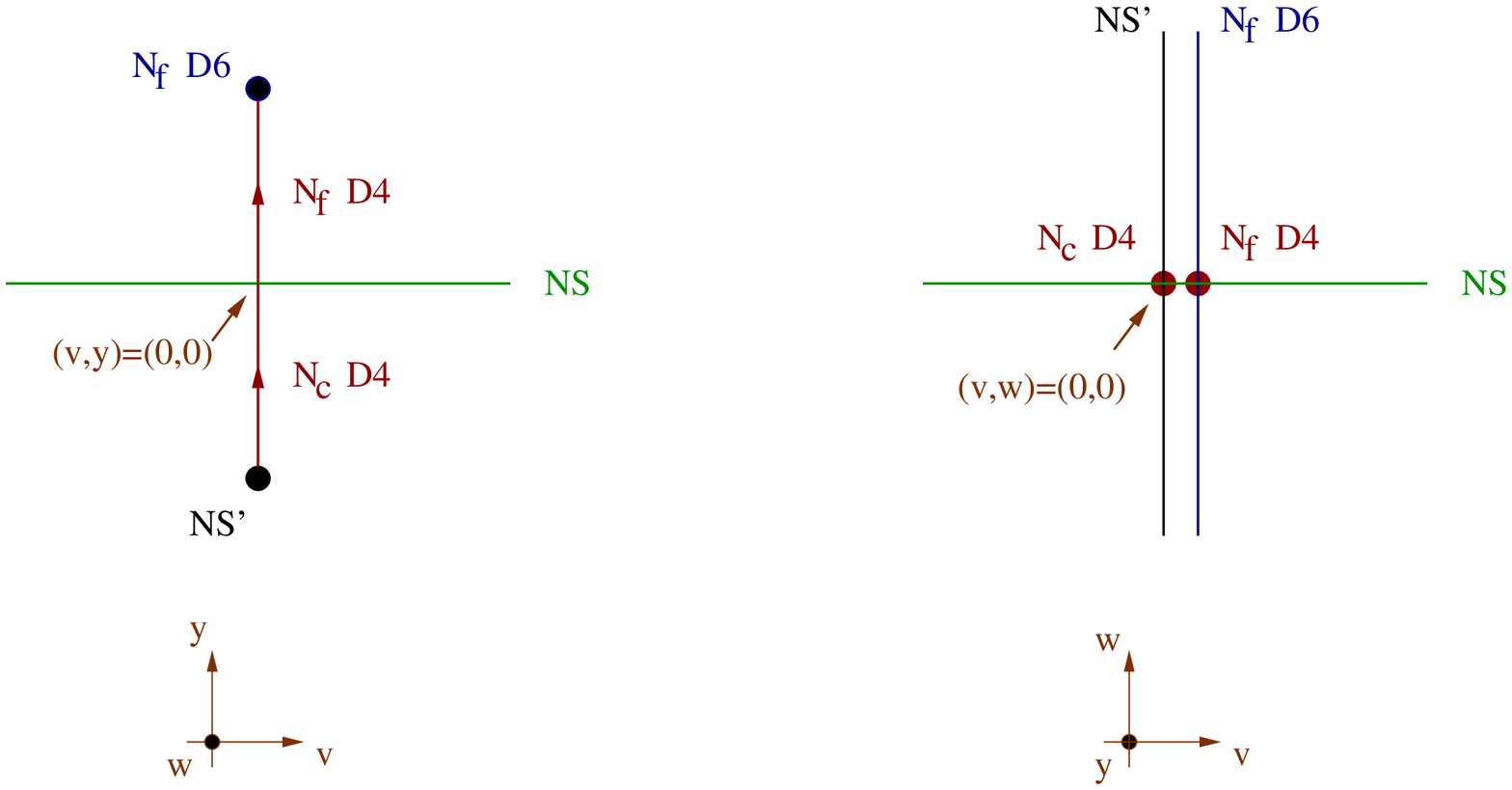}}

\noindent
It involves $N_c$ color $D4$-branes stretched between the $NS$ and $NS'$-branes,
which give rise to $U(N_c)$ SYM theory, and $N_f$ flavor $D4$-branes stretched
between the $D6$ and $NS$-branes, which give rise to $N_f$ flavors of quarks in
the fundamental representation of the gauge group, $Q_i$, $\widetilde Q^i$. This
construction, and in particular the relation between the configurations of figures
1 and 6, is discussed in \refs{\ElitzurFH,\ElitzurHC,\SchmaltzSQ,\HoriIW}
and reviewed in \GiveonSR.

Like in the magnetic description of the previous subsections, we can now translate
the $D6$-branes by an amount $v_2$ in the $(45)$ plane and rotate them by an angle
$\theta$ into the direction $w_\theta$ \strns. The resulting brane configuration is
described in figure 7, which is the electric analog of figure 3. Note that the two
figures are identical, with the replacements $v\leftrightarrow w$,
$\theta\leftrightarrow{\pi\over2}-\theta$, $N_c\leftrightarrow N_f-N_c$.

\ifig\loc{Supersymmetric vacua of the electric configuration.}
{\epsfxsize5.0in\epsfbox{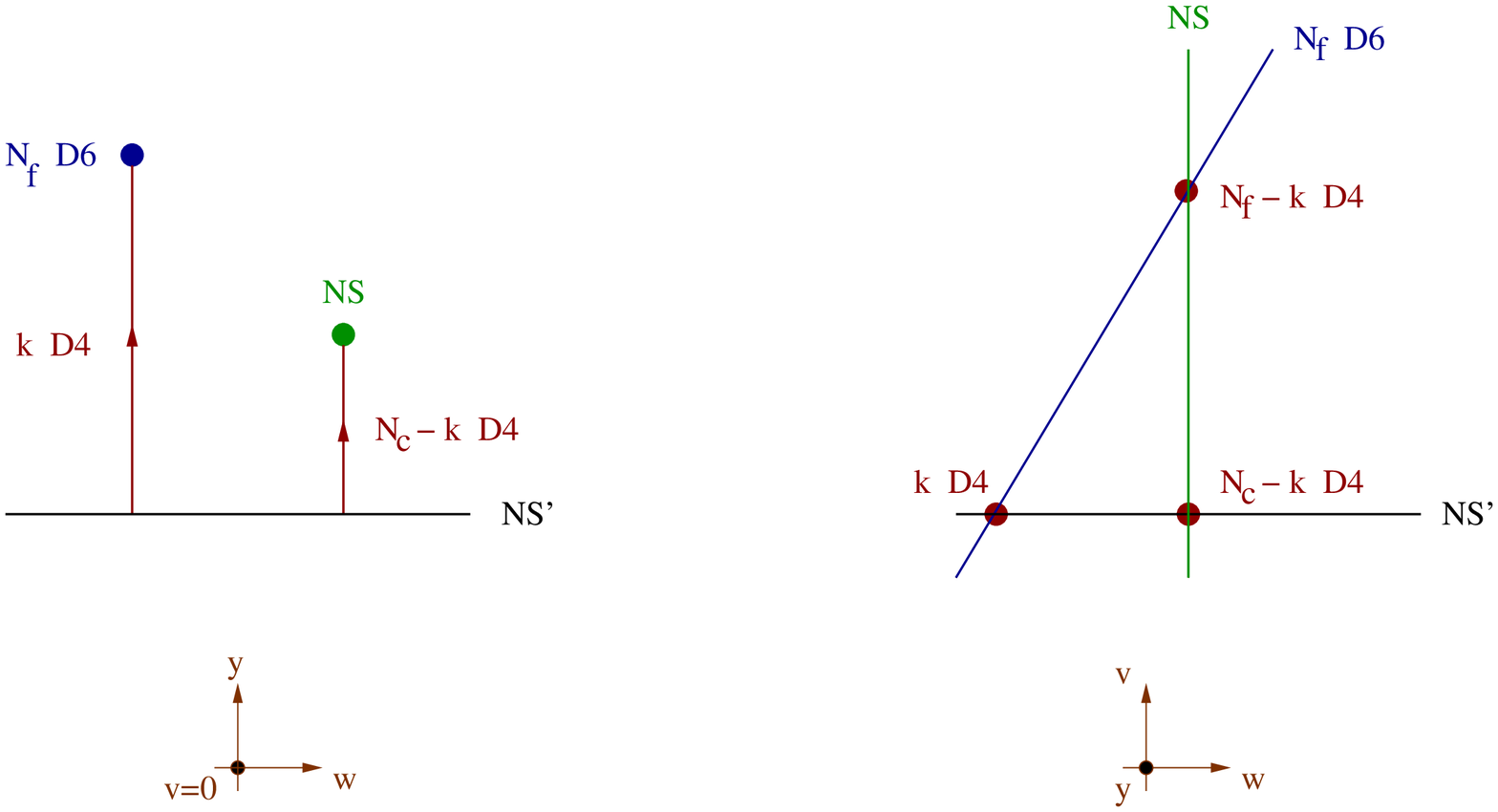}}

In the electric gauge theory the deformation of figure 7 corresponds to turning on a
superpotential for the quarks $Q$, $\widetilde Q$,
\eqn\wela{W_{\rm el}=\Tr\left[{\alpha\over 2}(\widetilde Q Q)^2-m\widetilde Q Q\right]=
\Tr\left({\alpha\over 2}M^2-mM\right)~.}
The parameters $m$ and $\alpha$ take the same values in terms of the geometric quantities
as before, \mmm\ and \altt, respectively. As in the magnetic analysis, one can replace the
superpotential \wela\ by \GiveonEF\
\eqn\weln{W_{\rm el}=-{1\over\Lambda}\widetilde Q^iN_i^jQ_j
-\Tr\left({\alpha_e\over 2}N^2-m_eN\right)~.}
The requirement that integrating out the massive fields $N_i^j$ gives \wela\
implies that
\eqn\aamm{\alpha\Lambda={1\over\alpha_e\Lambda}~,\qquad m={m_e\over\alpha_e\Lambda}~.}
The fields $N_i^j$ are the lightest mode of open strings ending on the flavor branes,
in complete analogy with the discussion of $M_i^j$ in subsection 2.1. They become
massless when $\alpha_e\to 0$, \ie\ $\theta\to{\pi\over2}$.

Classical supersymmetric vacua of the electric brane configuration are labeled by the
parameter $k$ indicated in figure 7, which runs over the range
\eqn\kele{k=0,1,\dots,N_c~.}
The unbroken gauge symmetry in the $k$'th configuration is $U(N_c-k)$. All this is
in agreement with the classical vacua found in  gauge theory  \GiveonEF.

\newsec{Metastable vacua}

In the previous section we described certain supersymmetric intersecting brane
configurations in type IIA string theory and compared them to the classical
ground states of  the low energy gauge theory on the branes. In \GiveonEF\ it was
shown that when quantum effects  are taken into account, additional non-supersymmetric
metastable vacua appear in this gauge theory, at least when $N_f<{3\over 2}N_c$ and the
coupling $\alpha$ is small. The existence of these vacua relies on a balance between
the classical and one loop contributions to the effective potential of the light fields.

In the brane realization, it was shown in \GiveonFK\ (in a closely related setting)  that
in the regime where the dynamics of the branes is well described by classical string
theory, the one loop field theory effects of \IntriligatorDD\ are replaced by the classical
gravitational attraction of the $D4$-branes to $NS5$-branes. Thus, we expect the
metastable vacua of \GiveonEF\ to appear in the systems described in section 2 when we
take this attraction into account. The purpose of this section is to show that this is indeed
the case.

It will turn out that the fivebrane whose gravitational potential plays a role
in our problem is the $NS$-brane \extbranes. In order to take its contribution into account
we have to study the motion of the $D4$-branes in the CHS geometry \CallanAT,
\eqn\chs{\eqalign{ &ds^2=dx_\mu dx^\mu+H(x^n) dx_mdx^m~,\cr
&e^{2(\Phi-\Phi_0)}=H(x^n)~,\cr
&H_{mnp}=-\epsilon_{mnp}^q\partial_q\Phi~.\cr }}
Here $\mu=0,1,2,3,4,5$; $m=6,7,8,9$; $H_{mnp}$ is the field strength
of the Neveu-Schwarz $B$ field; $g_s=\exp\Phi_0$ is the string
coupling far from the fivebranes. The harmonic function $H$ is given
by
\eqn\hcoin{H(r)=1+{l_s^2\over r^2}~,}
with $r^2=x_mx^m$. The background \chs\ is valid when $r\gg l_s$, and we will assume 
this throughout our discussion.

We will first consider a brane system in which the $D6$ and $NS'$-branes are separated
in $v$ and stretch in $w$, such as the configuration of figure 5 with
$\theta_1=\theta_2={\pi\over 2}$. This is the brane system studied in
\refs{\OoguriBG\FrancoHT-\BenaRG,\GiveonFK}.
In the flat space limit, $D4$-branes stretched between the $D6$ and $NS'$-branes in
this configuration can move in the $w$ direction without any cost of energy. These
are the brane analogs of the pseudo-moduli of \IntriligatorDD.

In the $NS$-fivebrane  geometry \chs, the pseudo-moduli acquire a mass \GiveonFK.
Indeed, if we hold the ends of the $D4$-branes at a fixed value of $w$, their energy
depends quadratically on this value (for small $w$).  We next calculate this energy as
a function of $w$  for the case $y_1=y_2$ in figure 5, and  comment on the case
$y_1\neq y_2$. Then we tilt the branes by changing $\theta_1, \theta_2$ and find a
locally stable equilibrium configuration. Finally,  we use the above results to describe
the pattern of locally stable magnetic brane configurations and match them to the gauge
theory analysis of \GiveonEF.

\subsec{Parallel $D6$ and $NS'$-branes}

In this subsection we consider the configuration of figure 5 with $\theta_1=\theta_2={\pi\over2}$.
Thus, we have $D4$-branes stretched between parallel $D6$ and $NS'$-branes and separated
by the distance
\eqn\ddeell{\Delta x=|v_1-v_2|}
along the $NS$-brane. In  \GiveonFK\ we calculated the energy density of such $D4$-branes;
here we would like to generalize the calculation to the case where the two ends of the fourbranes
are displaced to $w\not=0$.

\ifig\loc{A $D4$-brane displaced in $w$ is attracted to the $NS$-brane.}
{\epsfxsize2.3in\epsfbox{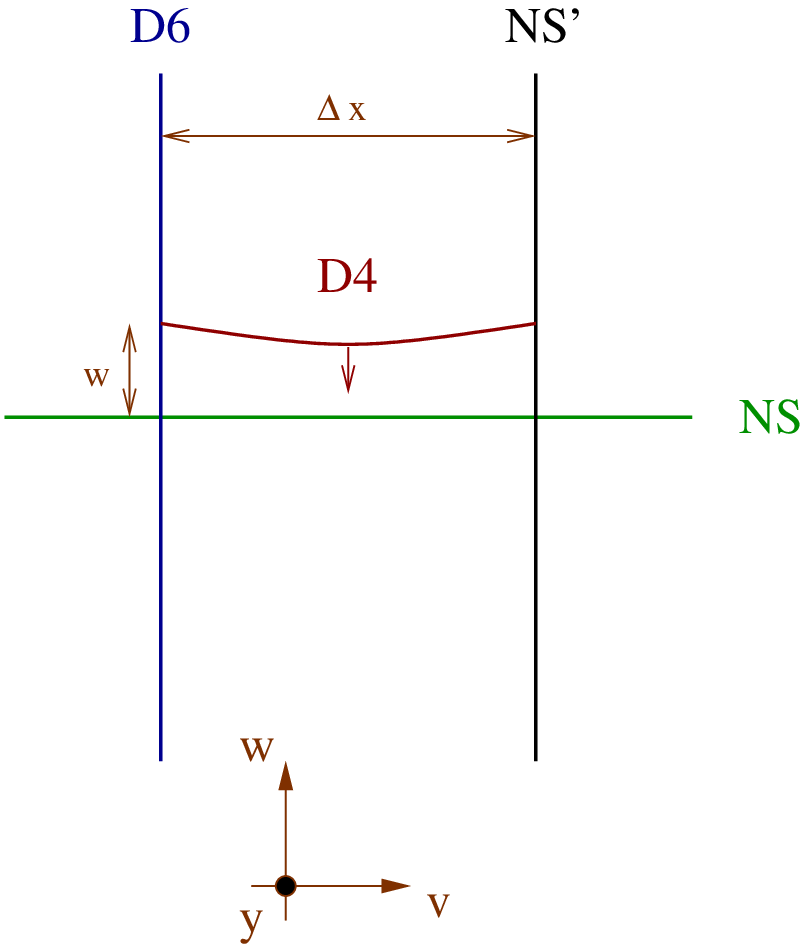}}

\noindent
To analyze this problem we consider the brane configuration in figure 8, in which a single $D4$-brane
is stretched between parallel $D6$ and $NS'$-branes, with its ends held fixed at an arbitrary value of 
$w$. We discuss in detail the case where the $D6$ and $NS'$-branes are at the same value of $y$, 
$y_1=y_2=y$, and then comment on the generalization to arbitrary $y_j$.

The energy of the $D4$-brane in figure 8 can be read off the results of \GiveonFK.  The only difference
between the present situation and the one there  is that the distance between the $NS$-brane and the
other extended branes, $y$, should be replaced by
\eqn\yyww{y_w=\sqrt{y^2+|w|^2}~.}
Remembering that we set the number of $NS$-branes $k$ to one,  equations (3.11), (3.12) in
\GiveonFK\ take the form
\eqn\ddxxyy{ \Delta x={y_w^2\over l_s}\sin2\theta_w+2l_s\theta_w~,}
and
\eqn\ddff{y_m=y_w\cos\theta_w~,}
where $y_m$ is the smallest value of $y$ along the $D4$-brane.

The energy density of the $D4$-brane is given by\foot{$\tau_4$ is the tension of a $D4$-brane
in flat spacetime.}
\eqn\curvede{E(w)=2\tau_4{y_my_w\over l_s}\sqrt{H(y_m)}\sin\theta_w~.}
It is convenient to rescale $E$ and define
\eqn\defetilde{\widetilde E\equiv{l_s\over2\tau_4} E~,}
which satisfies
\eqn\formtilde{\widetilde E^2=
y_w^2H(y_m)y_m^2\sin^2\theta_w=l_s^2y_w^2\sin^2\theta_w+
{1\over4}y_w^4\sin^22\theta_w~.}
Differentiating \formtilde\ with respect to $w$ we find
\eqn\formderiv{\partial_w\widetilde E^2=l_s^2\bar w\sin^2\theta_w+
l_s^2y_w^2(\sin2\theta_w)\partial_w\theta_w+\half y_w^2\bar
w\sin^22\theta_w +\half y_w^4(\sin4\theta_w)\partial_w\theta_w~.}
In order to finish the calculation we need to calculate $\partial_w\theta_w$.
This can be done by differentiating eq. \ddxxyy\ with respect to $w$, which leads to
\eqn\difftheta{(l_s^2+y_w^2\cos2\theta_w)\partial_w\theta_w+\half\bar
w\sin2\theta_w=0~.}
Equation \difftheta\ implies that the sum of the last three terms in \formderiv\
vanishes, so that
\eqn\finderive{\partial_w\widetilde E^2=l_s^2\bar w\sin^2\theta_w~.}
The only stationary point of \finderive\ is $w=0$. Indeed, if
$w\not=0$, \finderive\ only vanishes for $\theta_w=0, \pi/2$, and
these values are unphysical (for generic $\Delta x$) according to
\ddxxyy, \ddff. This is reasonable, since if we move the
$D4$-brane in $w$, the attraction to the $NS$-branes provides a
restoring force and we do not expect a stationary point at finite
$w$.

Expanding the energy \curvede\ around $w=0$, one finds that the
mode corresponding to displacement of the $D4$-brane in $w$ is
massive, as expected,
\eqn\quadord{E(w)={2\tau_4\over l_s}y\sqrt{l_s^2+y_m^2}\sin\theta_0+
\tau_4 l_s{\sin\theta_0\over y\sqrt{l_s^2+y_m^2}}w\bar w+O(|w|^4)~.}
In equation \quadord, $y_m$ is the smallest value of $y$ along the $D4$-brane
when the latter is placed at $w=0$, as in \GiveonFK, and $\theta_0$ is
the value of $\theta_w$ at $w=0$. Also, since this equation was obtained
from a supergravity analysis, it is valid in the regime $y\gg l_s$,  where it
can be simplified:
\eqn\sugraregime{E(w)=\tau_4\Delta x+{2\tau_4\over\Delta x}(\sin^2\theta_0)|w|^2+
O(|w|^4)\simeq \tau_4\Delta x+{\tau_4l_s^2\Delta x\over2y^4}|w|^2+
O(|w|^4)~.}
To calculate the mass of the mode $w$ we need to specify its kinetic term.
This can be done in the flat spacetime approximation, to which the fivebrane
geometry \chs\ only provides a small correction. In this approximation the
$D4$-brane is just a line segment of length $\Delta x$ located at a fixed $w$.
Thus, its kinetic term is given by the standard result
\eqn\kinww{\CL_k=-{\tau_4\over2}\Delta x|\partial_\mu w|^2~.}
Adding to this the potential energy density \sugraregime\ we find
that the mass of $w$ is
\eqn\masswww{m_w={l_s\over y^2}~.}
This mass is well below the string scale in the regime of validity of the supergravity
approximation. Interestingly, it does not depend on the separation of the branes $\Delta x$
\ddeell. If this behavior persisted to arbitrarily small $\Delta x$, this would be a problem
in our analysis below of the vacuum structure of the configuration of figure 5,
since as explained in \GiveonFK, in that regime gauge theory should take over,
and the leading contribution to the mass $m_w$ should come from the one loop effect
calculated in \IntriligatorDD. This effect leads to a mass $m_w^2\sim \Delta x$ which
goes to zero as $\Delta x\to 0$, and in particular is smaller than \masswww\ for sufficiently
small $\Delta x$.

The resolution of this is that in the brane construction of magnetic SQCD
it is important to take $y_2>y_1$ so that the configuration of figure 1 is 
non-singular. For $\Delta y=y_2-y_1\not=0$, the previous discussion is 
generalized in two important ways. The kinetic term \kinww\ involves the 
length of the $D4$-brane, which is now $\sqrt{(\Delta x)^2+(\Delta y)^2}$. 
Thus, in the limit $\Delta x\to 0$, the kinetic term approaches a finite limit,
\eqn\kinww{\CL_k=-{\tau_4\over2}\Delta y|\partial_\mu w|^2~.}
The second difference  involves the potential energy \sugraregime.
For $y_1=y_2$ the  coefficient of $|w|^2$ in the energy density $E$
went to zero like $\Delta x$; for $y_1\not= y_2$ we expect it to vanish
like $(\Delta x)^2$. The reason is that for $\Delta x=0$ the configuration
of figure 8 is supersymmetric for all $w$, and therefore the energy of the
$D4$-brane is independent of $w$. Furthermore, the limit $\Delta x\to 0$
is clearly non-singular and the dynamics is invariant under $\Delta x\to -\Delta x$,
so the energy should be analytic in $\Delta x$ near the origin.

Therefore, for $y_1\not= y_2$ we expect the analog of the mass \masswww\ to vanish like
\eqn\masssep{m_w\sim \Delta x}
as $\Delta x\to 0$. This contribution to the mass is smaller than the gauge theory one computed
in \IntriligatorDD, which goes like $\sqrt{\Delta x}$, as expected on general grounds.

\subsec{Non-parallel $D6$ and $NS'$-branes}

To analyze the system of figure 5 we need to generalize the discussion of the
previous subsection to configurations where the $D6$ and $NS'$-branes are
not parallel. To see what happens in these cases we turn to the configuration of
figure 9, where the branes are placed as follows in $\IC_v\times \IC_w$:
\eqn\shapensprime{\eqalign{D6:&\qquad\qquad v=-v_1+tw~,\cr
                           NS':&\qquad\qquad v=v_1-t w~,\cr
                           NS:&\qquad\qquad w=0~.\cr
}}
Thus, they are tilted by equal and opposite angles. We will continue to assume
that they are located at the same value of $y$. The generalization to non-equal
angles and $y_1\not=y_2$ is in principle straightforward.

\ifig\loc{A $D4$-brane stretched between tilted branes.}
{\epsfxsize2.3in\epsfbox{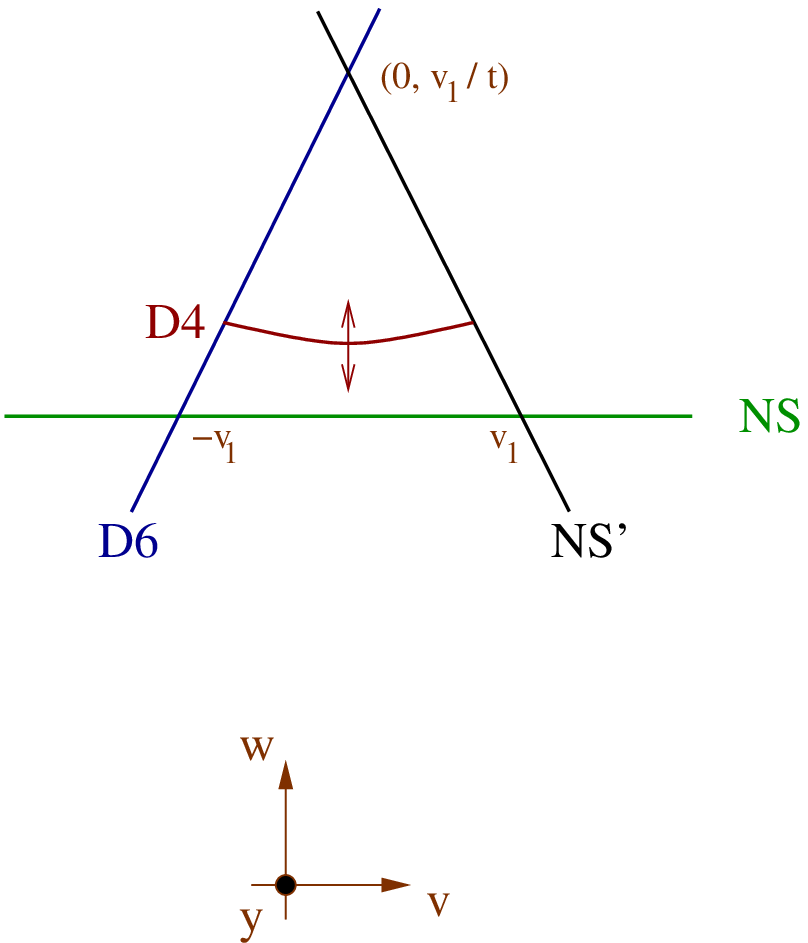}}

\noindent
The new element in  figure 9 is the dependence of the distance between the $D6$
and $NS'$-branes along the $NS$-branes, $\Delta x$ \ddeell,  on $w$:
\eqn\wdepdelx{\Delta x=2|v_1-tw|~.}
This leads to an extra contribution to $\partial_w\theta_w$ \difftheta,
which now takes the form
\eqn\newdiff{(l_s^2+y_w^2\cos2\theta_w)\partial_w\theta_w+\half\bar
w\sin2\theta_w -\half  l_st{\bar t\bar w-\bar
v_1\over|tw-v_1|}=0~.}
Plugging this into \formderiv\ we find
\eqn\finderive{\partial_w\widetilde E^2=l_s^2\bar
w\sin^2\theta_w+\half  l_sty_w^2\sin2\theta_w {\bar t\bar
w-\bar v_1\over|tw-v_1|}~.}
One place where this must (and does) vanish is at $w=v_1/t$, where
$\Delta x$ \wdepdelx, $\theta_w$ \ddxxyy\  and the energy density $E(w)$
\curvede\ vanish. As is clear from figure 9, this is the global minimum of the
energy, in which the $D4$-brane approaches the intersection of the $NS'$
and $D6$-branes.

Physically one would expect to find another local minimum of the energy,
at which the gravitational attraction of the $D4$-brane to the $NS$-brane is
precisely balanced by the force attracting the $D4$-brane to the intersection
of the $D6$ and $NS'$-branes. Without loss of generality we can take all the
parameters $v_1, w, t$ to be real and positive, and look for a solution with
\eqn\const{v_1>t w~.}
Setting the right hand side of \finderive\ to zero\ leads to
\eqn\consttw{\tan\theta_w={ty_w^2\over wl_s}~.}
To understand the form of the solution, let us look at the case
where gravity is weak. To be more precise, we will consider the
case where $v_1\gg l_s$ and
\eqn\weakgr{v_1l_s\ll y^2\le y_w^2~.}
To understand this inequality, recall that in \GiveonFK\ it was
shown that when $v_1l_s$ reaches a value of order $y^2$, a
$D4$-brane stretched between the $NS'$ and $D6$-branes becomes locally
unstable. The inequality \weakgr\ is the requirement that we stay
away from this regime of strong classical gravitational effects.

In this regime  $\theta_w$  \ddxxyy\ is small, so  \consttw\ takes the form
\eqn\formthw{\theta_w\simeq {ty_w^2\over wl_s}~.}
Plugging this into \ddxxyy\ one finds the location of the local minimum of
the effective potential,
\eqn\ddwwfin{w\simeq {ty_w^4\over l_s^2v_1}~.}
Plugging \ddwwfin\ back into \formthw\ one can check that
in the regime \weakgr\ $\theta_w$ is indeed very small.

Note that in deriving \ddwwfin\ we assumed that
\eqn\alpww{t w\ll v_1~,}
so we can approximate $\Delta x\simeq 2v_1$ in \wdepdelx. In order
for this to be the case, it must be that
\eqn\boundalp{t\ll {v_1l_s\over y_w^2}\ll 1~.}
As is clear from figure 9, this implies that the angle between the $D6$ and $NS'$-branes
is very small. It is not difficult to generalize the discussion to cases where that angle is
not small.

One might also want to require that the local minimum occurs at a
value of $w$ much smaller than $y$. In that case one can replace
$y_w$ by $y$ in \ddwwfin. This leads to
\eqn\ddwwsmall{w\simeq t{y^4\over l_s^2v_1}~.}
The requirement $w\ll y$ implies
\eqn\smallalpha{t\ll {v_1l_s^2\over y^3}~,}
a more stringent bound than \boundalp.

To summarize, we find that, as one would expect, the brane configuration of figure 9
has a local minimum where the ends of the $D4$-brane are located at $w$ given by
\consttw. When the slope parameter $t$ is small \smallalpha, the minimum is located
at a small value of $w$ \ddwwsmall.

\subsec{Metastable states in the brane construction of SQCD}

In the previous subsection we saw that a $D4$-brane stretched between tilted
$NS'$ and $D6$-branes, as in figure 9, has a locally stable configuration in which
its ends are at a non-zero value of $w$ given by \consttw, \ddwwsmall. In this
subsection we would like to explore the implications of this for the
brane realizations of SQCD discussed in section 2.

For concreteness we will restrict to the magnetic brane configuration depicted in figure 3.
In the $k$'th supersymmetric vacuum there are $N_f-k$ $D4$-branes stretched between
the $NS'$ and $D6$-branes at $w=-v_2\cot\theta$.  We can move $n$ of these $D4$-branes
towards the $NS$-brane to the local minimum of the potential found in the previous
subsection. As we saw there, for small $\theta$ this minimum occurs at a small value of
$w$ \ddwwsmall.  The resulting brane configuration is shown in figure 10.

\ifig\loc{Metastable vacua of the deformed magnetic configuration.}
{\epsfxsize5.0in\epsfbox{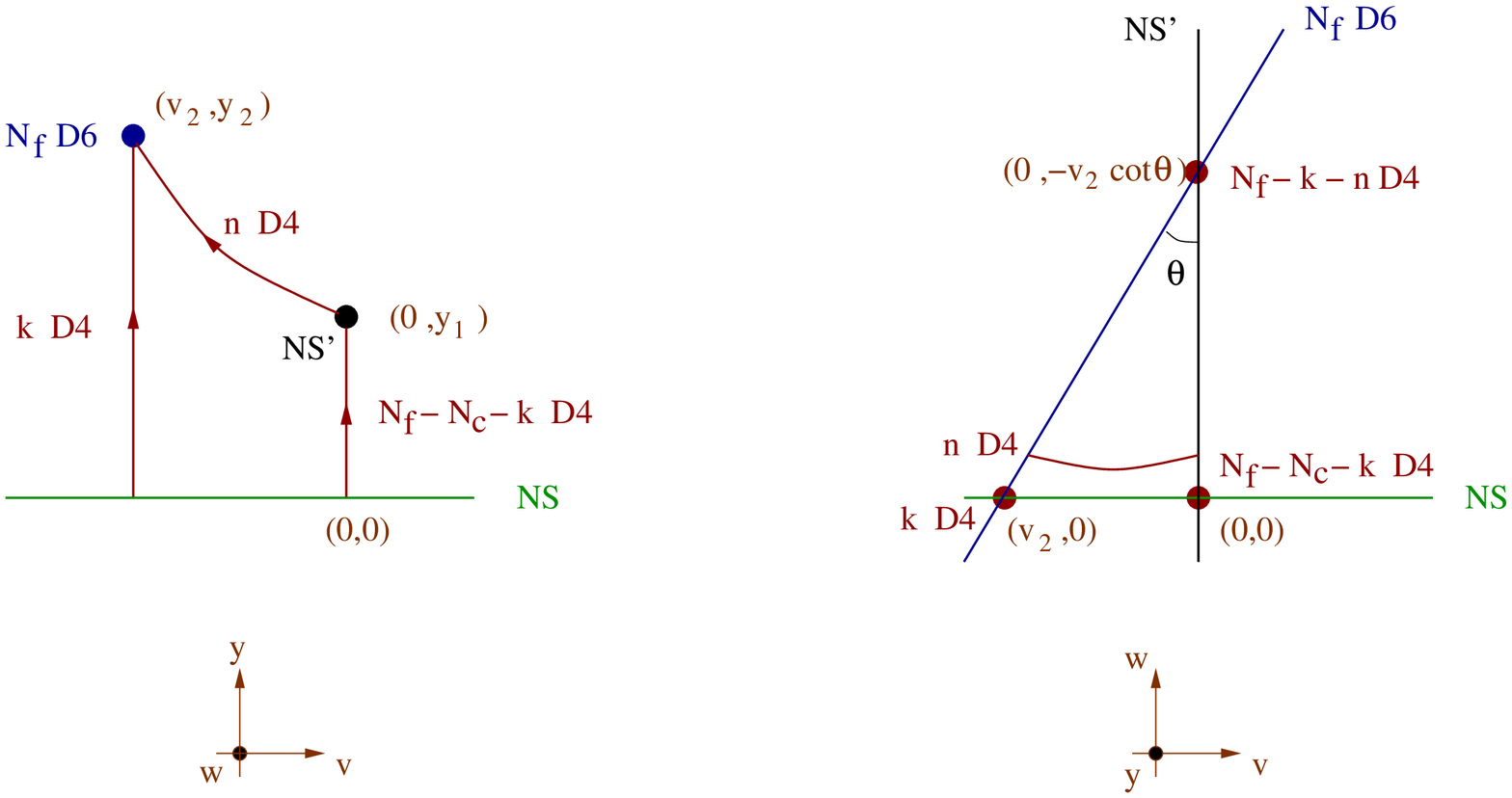}}

The left figure shows the vicinity of $w=0$ in the brane configuration. The endpoints of the
$n$ $D4$-branes stretched between the $NS'$ and $D6$-branes are at a non-zero $w$, as
can be seen in the right figure.

While the brane configuration of figure 10 is locally stable, it can decay to the supersymmetric
ground states of section 2. For $N_f-N_c-k>0$ there are two types of instabilities.
One involves a process where  the endpoints of the $n$ flavor $D4$-branes on the $NS'$-brane
approach those of the $N_f-N_c-k$ color ones, the two types of branes connect and  move to the
intersection of  the $D6$ and $NS$-branes. For $n\le N_f-N_c-k$, the endpoint of this process 
is the supersymmetric vacuum of figure 3 with $k\to k+n$. Otherwise, some of the $n$ flavor 
$D4$-branes remain.

\noindent
A second instability has the $n$ $D4$-branes moving to larger $w$, back to the configuration
of figure 3.  Of course, one can also consider processes where some of the $D4$-branes
approach the $NS'$ -- $D6$ intersection while the rest approach the $NS$ -- $D6$ one. For
$N_f-N_c-k=0$, the first type of instability is absent. Therefore, such vacua are more long-lived.
More generally, this is the case for $N_f-N_c-k<n$. 

In all the processes discussed above, the energy of the $D4$-branes first increases and then
decreases to the supersymmetric value. Thus, these are non-perturbative instabilities whose
resolution involves tunneling. For fixed values of all the geometric parameters in the limit $g_s\to 0$,
the lifetime of the metastable states goes like $\exp(C/g_s)$. The constant $C$ depends on
the particular state and would be interesting to calculate.

The metastable states of figure 10 are in one to one correspondence  with those
constructed in the magnetic gauge theory in section 3 of \GiveonEF. In that analysis,
the effective potential of the magnetic gauge theory was found to have local minima
in which the fields $M$ and $\tilde qq$ have the form
\eqn\metaone{M=\left(\matrix{0&0&0\cr 0&M_n&0\cr
0&0&{m\over\alpha}I_{N_f-k-n}\cr}\right)~,}
and
\eqn\metatwo{\tilde qq=\left(\matrix{m\Lambda I_k&0&0\cr
0&0&0\cr 0&0&0}\right)~,}
where the $n\times n$ matrix $M_n$ is given (up to an order one factor) by
\eqn\mmnn{M_n\simeq{\alpha\Lambda^3\over N_f-N_c}I_n~.}
These metastable states correspond to the ones in figure 10, with the parameters
$k$ and $n$ that appear in both identified.

In fact, the gauge theory analysis of \GiveonEF\ is directly applicable to the brane
configurations constructed here in a particular regime in their parameter space.
In gauge theory it is natural to write the superpotential \wmagmb\ as
\eqn\wwmm{W_{\rm mag}=h\tilde q_i\Phi^i_j q^j- {\rm Tr}\left(h\mu^2\Phi-\half
h^2\mu_\phi\Phi^2\right)}
where $\Phi$ is proportional to $M$ and has a canonical kinetic term \GiveonEF.
The couplings $h$, $\mu$, $\mu_\phi$ are given (up to order one factors) in
terms of the brane parameters by (see \altt\ and \BenaRG)
\eqn\hmuphi{h^2={g_sl_s\over y_2-y_1},\qquad \mu^2=-{v_2\over g_s l_s^3},
\qquad \mu_\phi={\tan\theta\over g_s l_s}~.}
The analysis of \GiveonEF\ is valid for
\eqn\constmmuu{\mu_\phi\ll\mu\ll m_s~, \;\;\;\;\;h\ll 1~,}
such that the physics is perturbative in $h$, $\mu_\phi/\mu$, and takes place well
below the string scale.  Plugging in the values  \hmuphi\ leads to the constraints
\eqn\ccll{\tan^2\theta\ll {v_2g_s\over l_s}\ll g_s^2\ll \left(y_2-y_1\over l_s\right)^2~.}
Thus, the field theory analysis is valid when $\theta$ and $v_2/l_s$ are much smaller
than $g_s$.

The classical brane construction generalizes the field theory discussion to the regime
where the angle $\theta$ is of order one and the different length parameters in
figure 3 are of order $l_s$ or larger. In this regime the gauge theory analysis is not
valid, but as we see the phase structure is essentially identical.

All the elements of the gauge theory discussion have direct analogs in the brane construction:
\item{(1)} The $n$ light flavors of $SU(N_f-N_c-k)$, denoted by $\varphi$, $\tilde\varphi$ in \GiveonEF,
correspond in the brane picture of figure 10 to fundamental strings stretched between the
$n$ flavor $D4$-branes and the $N_f-N_c-k$ color ones.
\item{(2)} The tachyonic instability found in gauge theory for $\mu_\phi<h\mu$ (see eq. (3.21) in
\GiveonEF) is due in the brane construction to the fact that the ground state of the above fundamental
strings is tachyonic when the angle $\theta$ and thus the distance between the endpoints of the
color and flavor branes along the $NS'$-brane is sufficiently small.
\item{(3)} The one loop effects that are necessary for stabilizing the metastable states in \GiveonEF\
are replaced in the brane picture by the classical gravitational attraction of the $D4$-branes to the
$NS$-brane, as in \GiveonFK.

\noindent
An interesting limit of the brane configuration of figure 10 is $\theta\to 0$ with the intersection
point of the $NS$ and $D6$-branes $(v_2,0)$ held fixed. The resulting configuration describes SQCD
with the superpotential $W=-mM$ \refs{\OoguriBG\FrancoHT-\BenaRG,\GiveonFK}. It is natural to ask
what happens to all the metastable states described above in this limit. The states with
$1\le n\le N_f-N_c-k$  become perturbatively unstable below a critical value of $\theta$, as mentioned
in point (2) above. Condensation of the tachyon mentioned there leads to a supersymmetric vacuum of 
the sort depicted in figure 3, with $k\to k+n$. For $n>N_f-N_c-k$, tachyon condensation leaves some
flavor branes that cannot decay in this way.
These appear to give rise to metastable states in the theory with $\theta=0$.

One can exhibit all such vacua by taking $N_f-N_c-k=0$, and letting $n$ run over the range
$n=1,\cdots, N_c$. For $n=N_c$ this procedure leads to the states considered in \IntriligatorDD.
For $0<n<N_c$ one finds additional states not considered in \IntriligatorDD. These states have the
property that as $\theta\to 0$, $N_f-k-n$ $D4$-branes in figure 10 go to infinity in $w$. One expects
quantum corrections to modify the physics of these fourbranes, but since their dynamics takes place
far from the $n$ flavor $D4$-branes that give rise to the metastable states, it is not clear that
these effects should influence the metastable states. This is certainly the case in the gravity regime,
where figure 10 is reliable, and it would be interesting to see whether such states exist in gauge
theory as well.

\newsec{Discussion}

The fact that intersecting NS and D-brane constructions of the sort reviewed in \GiveonSR\
provide a useful guide for the analysis of supersymmetric ground states in various quantum
field theories has been known for some time. The main conclusion of the present investigation
is that this is the case for metastable non-supersymmetric ground states as well. We found that
taking into account the gravitational attraction of the D-branes to the NS-fivebranes leads to
a rich landscape of metastable states which are very similar to the corresponding gauge theory
ground states.

This construction can be generalized in many ways discussed in the supersymmetric
context in the past \GiveonSR. In particular,
\item{(a)} Replacing the $N_f$ $D6_\theta$-branes by an $NS_\theta$-brane corresponds
in the low energy theory to gauging the $U(N_f)$ global symmetry.
\item{(b)} Increasing the number of  Neveu-Schwarz fivebranes leads to higher order polynomial
superpotentials for the chiral superfields in the adjoint of $U(N_c)\times U(N_f)$. For instance,
replacing the $NS_\theta$-brane of point (a) by $n_0$ coincident $NS_\theta$-branes and
separating them in the transverse direction, leads to a superpotential of the form
$W(M)=\sum_{n=1}^{n_0} \lambda_n\Tr M^n$ for  the adjoint of $U(N_f)$, $M$.
\item{(c)} Replacing the $NS'$-brane in figure 1 (and subsequent figures) with a second stack of
$D6$-branes leads instead to an O'Raifeartaigh-type model with no gauge fields,
of the type studied recently in \IntriligatorPY.

\noindent
These and other generalizations of the construction of this paper can be analyzed along
the same lines, and presumably lead to a rich structure. It might be interesting to use
such constructions to embed models of gauge mediation and their stringy generalizations
in string theory.

\bigskip
\noindent{\bf Acknowledgements:}
This work is supported in part by the BSF -- American-Israel
Bi-National Science Foundation. AG is supported in part by a
center of excellence supported by the Israel Science Foundation
(grant number 1468/06), EU grant MRTN-CT-2004-512194, DIP
grant H.52, and the Einstein Center at the Hebrew University. DK
is supported in part by DOE grant DE-FG02-90ER40560 and the
National Science Foundation under Grant 0529954. AG and DK
thank the EFI at the University of Chicago and Hebrew University,
respectively, for hospitality.

\listrefs
\end

Actually, in the regime where the DBI analysis is reliable, the
magnetic theory is weakly coupled for {\it any}
$N_f>N_c$. In the DBI regime -- $y_{1,2},\Delta x>l_s$, $g_s^2\ll 1$
-- when $N_f<{3\over 2}N_c$ one finds $\Lambda_m\gg m$, where $\Lambda_m$
is the dynamically generated scale of the magnetic SQCD, while
$\Lambda_m\ll m$ if $N_f>{3\over 2}N_c$ (see eq. (4.9) in
\GiveonFK). Hence, in both cases, the gauge theory is weakly
coupled at $E\simeq m$: in the first case the theory is IR free
and is thus weakly coupled at $E\ll\Lambda_m$, while in the second
case this magnetic theory is asymptotically free and thus weakly
coupled at $E\gg\Lambda_m$.\foot{Note that in the latter case
$E\gg\Lambda_m$ and thus the magnetic theory does {\it not} have
an electric Seiberg dual.} Consequently, the K\"ahler potential is
computable to sufficient approximation, and one finds long lived
meta-stable vacua for any $N_f>N_c$ also in the low energy gauge
theory: the lifetime behaves like $(\Lambda_m/m)^{C(3N_c-2N_f)}$,
where $C$ is some positive number, and is thus very large for any
$N_f>N_c$. This is strengthening the observation regarding the
gauge/gravity correspondence in our intersecting branes systems:
the DBI analysis, done in the gravity regime, produces the same
phenomena as seen in the -- apparently very different -- gauge
theory regime.

\lref\GiveonFK{
  A.~Giveon and D.~Kutasov,
  ``Gauge symmetry and supersymmetry breaking from intersecting branes,''
  Nucl.\ Phys.\  B {\bf 778}, 129 (2007)
  [arXiv:hep-th/0703135].
}

\lref\AharonyTI{
  O.~Aharony, S.~S.~Gubser, J.~M.~Maldacena, H.~Ooguri and Y.~Oz,
  ``Large N field theories, string theory and gravity,''
  Phys.\ Rept.\  {\bf 323}, 183 (2000)
  [arXiv:hep-th/9905111].
}

\lref\OoguriPJ{
  H.~Ooguri and Y.~Ookouchi,
  ``Landscape of supersymmetry breaking vacua in geometrically realized gauge
  theories,''
  Nucl.\ Phys.\ B {\bf 755}, 239 (2006)
  [arXiv:hep-th/0606061].
}

\lref\ArgurioNY{
  R.~Argurio, M.~Bertolini, S.~Franco and S.~Kachru,
  ``Gauge / gravity duality and meta-stable dynamical supersymmetry breaking,''
  arXiv:hep-th/0610212.
}

\lref\AganagicEX{
  M.~Aganagic, C.~Beem, J.~Seo and C.~Vafa,
  ``Geometrically induced metastability and holography,''
  arXiv:hep-th/0610249.
}

\lref\HeckmanWK{
  J.~J.~Heckman, J.~Seo and C.~Vafa,
  ``Phase Structure of a Brane/Anti-Brane System at Large N,''
  arXiv:hep-th/0702077.
}

\lref\OoguriWJ{
  H.~Ooguri and C.~Vafa,
  ``Two-Dimensional Black Hole and Singularities of CY Manifolds,''
  Nucl.\ Phys.\ B {\bf 463}, 55 (1996)
  [arXiv:hep-th/9511164].
}

\lref\KutasovTE{
  D.~Kutasov,
  ``Orbifolds and Solitons,''
  Phys.\ Lett.\ B {\bf 383}, 48 (1996)
  [arXiv:hep-th/9512145].
}

\lref\GiveonZM{
  A.~Giveon, D.~Kutasov and O.~Pelc,
  ``Holography for non-critical superstrings,''
  JHEP {\bf 9910}, 035 (1999)
  [arXiv:hep-th/9907178].
}

\lref\AhnGN{
  C.~Ahn,
  ``Brane configurations for nonsupersymmetric meta-stable vacua in SQCD with
  adjoint matter,''
  arXiv:hep-th/0608160.
}

\lref\TatarDM{
  R.~Tatar and B.~Wetenhall,
  ``Metastable vacua, geometrical engineering and MQCD transitions,''
  arXiv:hep-th/0611303.
}

\lref\IntriligatorDD{
  K.~Intriligator, N.~Seiberg and D.~Shih,
  ``Dynamical SUSY breaking in meta-stable vacua,''
  JHEP {\bf 0604}, 021 (2006)
  [arXiv:hep-th/0602239].
}

\lref\SakaiCN{
  T.~Sakai and S.~Sugimoto,
  ``Low energy hadron physics in holographic QCD,''
  Prog.\ Theor.\ Phys.\  {\bf 113}, 843 (2005)
  [arXiv:hep-th/0412141].
}

\lref\AntonyanVW{
  E.~Antonyan, J.~A.~Harvey, S.~Jensen and D.~Kutasov,
  ``NJL and QCD from string theory,''
  arXiv:hep-th/0604017.
}

\lref\AntonyanQY{
  E.~Antonyan, J.~A.~Harvey and D.~Kutasov,
  ``The Gross-Neveu model from string theory,''
  arXiv:hep-th/0608149.
}

\lref\AntonyanPG{
  E.~Antonyan, J.~A.~Harvey and D.~Kutasov,
  ``Chiral symmetry breaking from intersecting D-branes,''
  arXiv:hep-th/0608177.
}

\lref\CallanAT{
  C.~G.~Callan, J.~A.~Harvey and A.~Strominger,
  ``Supersymmetric string solitons,''
  arXiv:hep-th/9112030.
}

\lref\AharonyUB{
  O.~Aharony, M.~Berkooz, D.~Kutasov and N.~Seiberg,
  ``Linear dilatons, NS5-branes and holography,''
  JHEP {\bf 9810}, 004 (1998)
  [arXiv:hep-th/9808149].
}

\lref\AharonyXN{
  O.~Aharony, A.~Giveon and D.~Kutasov,
  ``LSZ in LST,''
  Nucl.\ Phys.\ B {\bf 691}, 3 (2004)
  [arXiv:hep-th/0404016].
}

\lref\KutasovDJ{
  D.~Kutasov,
  ``D-brane dynamics near NS5-branes,''
  arXiv:hep-th/0405058.
}

\lref\KitanoXG{
  R.~Kitano, H.~Ooguri and Y.~Ookouchi,
  ``Direct mediation of meta-stable supersymmetry breaking,''
  arXiv:hep-ph/0612139.
}

\lref\KutasovRR{
  D.~Kutasov,
  ``Accelerating branes and the string / black hole transition,''
  arXiv:hep-th/0509170.
}

\lref\ItzhakiZR{
  N.~Itzhaki, D.~Kutasov and N.~Seiberg,
  ``Non-supersymmetric deformations of non-critical superstrings,''
  JHEP {\bf 0512}, 035 (2005)
  [arXiv:hep-th/0510087].
}

\lref\LukyanovNJ{
  S.~L.~Lukyanov, E.~S.~Vitchev and A.~B.~Zamolodchikov,
  ``Integrable model of boundary interaction: The paperclip,''
  Nucl.\ Phys.\ B {\bf 683}, 423 (2004)
  [arXiv:hep-th/0312168].
}

\lref\LukyanovBF{
  S.~L.~Lukyanov and A.~B.~Zamolodchikov,
  ``Dual form of the paperclip model,''
  Nucl.\ Phys.\ B {\bf 744}, 295 (2006)
  [arXiv:hep-th/0510145].
}

\lref\KutasovCT{
  D.~Kutasov,
  ``A geometric interpretation of the open string tachyon,''
  arXiv:hep-th/0408073.
}

\lref\GiveonPR{
  A.~Giveon and D.~Kutasov,
  ``Fundamental strings and black holes,''
  arXiv:hep-th/0611062.
}

\lref\WittenSC{
  E.~Witten,
  ``Solutions of four-dimensional field theories via M-theory,''
  Nucl.\ Phys.\  B {\bf 500}, 3 (1997)
  [arXiv:hep-th/9703166].
}

\lref\AdamsSV{
  A.~Adams, J.~Polchinski and E.~Silverstein,
  ``Don't panic! Closed string tachyons in ALE space-times,''
  JHEP {\bf 0110}, 029 (2001)
  [arXiv:hep-th/0108075].
}

\lref\HarveyWM{
  J.~A.~Harvey, D.~Kutasov, E.~J.~Martinec and G.~W.~Moore,
  ``Localized tachyons and RG flows,''
  arXiv:hep-th/0111154.
}

\lref\NakayamaYX{
Y.~Nakayama, Y.~Sugawara and H.~Takayanagi,
``Boundary states for the rolling D-branes in NS5 background,''
JHEP {\bf 0407}, 020 (2004)
[arXiv:hep-th/0406173].
}

The scale $\Lambda$ can be read, e.g., from the relation
$\mu^2=m\Lambda$ and $\mu^2={v_1-v_2\over g_sl_s^3}$ \refs{\BenaRG,\GiveonFK,\GiveonEF}
(see e.g. eq. (3.5) in \GiveonEF, and eq. (2.11) in \GiveonFK).